\documentclass[iop]{emulateapj}
\usepackage[demo]{rotating}

\usepackage{epsfig}
\usepackage{amsmath}
\usepackage{natbib}
\usepackage{graphicx,color}
\usepackage{xcolor}
\usepackage{float}
\usepackage{mathtools}

\begin{document}

\title{Ground-based Spectroscopy of the exoplanet XO-2b using a Systematic Wavelength Calibration}

\author{Kyle A. Pearson}
\affil{ Lunar and Planetary Laboratory, University of Arizona, 1629 East University Boulevard, Tucson, AZ, 85721, USA }
\affil{ Department of Physics and Astronomy, Northern Arizona University, San Francisco Street, Flagstaff, AZ, 86001, USA }

\author{Caitlin A. Griffith}
\affil{ Lunar and Planetary Laboratory, University of Arizona, 1629 East University Boulevard, Tucson, AZ, 85721, USA }

\author{Robert T. Zellem}
\affil{ Jet Propulsion Laboratory, California Institute of Technology, 4800 Oak Grove Drive, Pasadena, CA, 91109, USA }

\author{Tommi T. Koskinen}
\affil{ Lunar and Planetary Laboratory, University of Arizona, 1629 East University Boulevard, Tucson, AZ, 85721, USA }

\author{Gael M. Roudier}
\affil{ Jet Propulsion Laboratory, California Institute of Technology, 4800 Oak Grove Drive, Pasadena, CA, 91109, USA  }

\begin{abstract}

Exoplanets orbiting close to their host star are expected to support a large ionosphere, which extends to larger pressures than witnessed in our Solar System. These ionospheres can be investigated with ground-based transit observations of the optical signatures of alkali metals, which are the source of the ions. However, \textcolor{black}{most} ground-based transit spectra do not systematically resolve the wings of the features and continuum, as needed to constrain the alkali abundances. Here, we present new observations and analyses of optical transit spectra that cover the Na doublet in the atmosphere of the exoplanet XO-2 b. To assess the consistency of our results, observations were obtained from two separate platforms: Gemini/GMOS and Mayall/KOSMOS. To mitigate the systematic errors, we chose XO-2, because it has a binary companion of the same brightness and stellar type, which provides an ideal reference star to model Earth's atmospheric effects. We find that interpretation of the data is highly sensitive to time-varying translations along the detector, which change according to wavelength and differ between the target and reference star. It was necessary to employ a time-dependent cross-correlation to align our wavelength bins and correct for atmospheric differential refraction. This approach allows us to resolve the wings of the Na line across 5 wavelength bins at a resolution of $\sim$1.6nm and limit the abundance of Na. We obtain consistent results from each telescope with a Na amplitude of 521$\pm$161 ppm and 403$\pm$186 ppm for GMOS and KOSMOS respectively. The results are analyzed with a radiative transfer model that includes the effects of ionization. The data are consistent with a clear atmosphere between $\sim$1--100 mbar which establish a lower limit on Na at 0.4$^{+2}_{-0.3}$ ppm \textcolor{black}{([Na/H]=-0.64$^{+0.78}_{-0.6}$)}, consistent with solar. However, we can not rule out the presence of clouds at $\sim$10 mbar which allow for higher Na abundances which would be consistent with stellar metallicity measured for the host star \textcolor{black}{([Na/H]=0.485$\pm$0.043)}.

\end{abstract}


\section{Introduction}

Transiting exoplanets enable studies of planetary atmospheres distinct from those in the Solar System. During primary transit, host star light is transmit through the planet's atmosphere thereby revealing absorption features from atomic and molecular species. The short orbital periods and large sizes of hot Jupiters enable ground and space-based measurements of their radii and atmospheric compositions. The most prominent features in optical spectra of hot-Jupiters from 1000-17000 K are Na, K and Rayleigh scattering \citep{Fortney2010}. 

Alkali metals are readily ionized in the hot atmospheres of close-in planets, producing an extensive ionosphere more like that of a star rather than of Jupiter. Atomic sodium and potassium have been detected on a number of hot Jupiters, e.g., HD 209458b (\citealt{Charbonneau2002}; \citealt{Snellen2008}), HD 189733b \citep{Redfield2008}, WASP-6 b \citep{Nikolov2015}, WASP-17b \citep{Wood2011}, XO-2N b \citep{Sing2011}, WASP-39 b \citep{Nikolov2016}, and HAT-P-1b \citep{Nikolov2014}. Atomic sodium and potassium produce optical doublet spectroscopic lines at 589.3~nm and 766.4~nm, respectively, which arise from transitions between their ground state and first excited state. Detections of Na and K indicate the depth of the ionosphere and source of electrons in the atmosphere (\citealt{Lavvas2014}; \citealt{Koskinen2014}). In the hot-Jupiter XO-2b (T$_{eq}$$\sim$1300K) Na and K are expected to remain in the gas phase since their condensation temperatures are 1156 K and 932 K, respectively \citep{Morley2012}. The condensation temperature of Na and K are similar, and both the Na I and K I doublet features are predicted in the spectra of warm exoplanets, even when considering a range of possible elemental abundances. 

Measurements of Na and K alkali features have led to puzzling results. Some exoplanets (e.g., XO-2b, WASP-39 b and WASP-6 b) indicate the presence of both Na and K, as predicted, while other studies of hot-Jupiters indicate only K I or only Na I (e.g., HAT-P-1b, WASP-17 b, WASP-31 b, HAT-P-12 b and HD 189733 b) \citep{Sing2016}. While space-based observations of HD 209458 b \citep{Sing2008} have defined the continuum and wings of the Na line, \textcolor{black}{ most transit spectra from the ground only resolve the narrow line core. Recently there have been some exceptions, ground-based measurements for the targets WASP-96 b and WASP-127 b exhibit enough definition in the line shape to place constraints on the alkali abundances (\citep{Nikolov2018}; \citep{Chen2018}).} In order to determine the Na abundance, measurements of the alkali band wings and the continuum are needed to define the spectral feature. Such spectral definition has been difficult to achieve particularly from ground-based observations potentially due to cloud and haze opacity on the day night terminator or lack of resolution and signal (e.g. \citealt{Sing2011}, \citealt{Sing2012}, \citealt{Wilson2015}, \citealt{Sedaghati2016}, \citealt{Gibson2017}). Constraints on Na and K abundances also require knowledge of the planet's radius at which the atmosphere becomes opaque to limit the degeneracies between retrieved abundances and the planet's opaque radius (\citealt{Tinetti2010}; \citealt{Benneke2012}; \citealt{Benneke2013}; \citealt{Griffith2014}). The presence of clouds can shape the continuum at optical wavelengths and complicate the detection of alkali band wings by dampening or masking the Na and K features (e.g. \citealt{Sing2016})

Our work investigates the ionosphere of hot Jupiter exoplanet with visible wavelength absorption spectroscopy of XO-2b. This target is chosen because the host star, XO-2N (G9V; $V$-mag=11.138), has a binary companion star (separated by 4600 AU = 31.1";\cite{Burke2007}), XO-2S, of similar stellar type (G9V) and brightness ($V$-mag = 11.086). Previous observations of XO-2b at optical wavelengths from the Gran Telescopio Canarias using a low resolution spectrograph and narrow band filters detected the presence of both Na and K, respectively (\citealt{Sing2011}; \citealt{Sing2012}). However, the abundances of Na and K were not constrained due to a lack of definition in the band wings, possibly due to scattering from clouds or hazes and instrumental systematics (i.e. seeing induced slit losses). Hubble Space Telescope NICMOS observations at 1.2--1.8~$\mu$m detected the tentative (1.75 $\sigma$) presence of water vapor in the atmosphere of XO-2b \citep{Crouzet2012}. This low spectral modulation due to water is degenerate with a low mixing ratio and$/$or the presence of clouds.

This paper presents and analyzes two nights of spectroscopic data, recorded by the visible-wavelength multi-object spectrographs of GMOS at the 8.2 meter Gemini Observatory and of KOSMOS at the 4m Mayall telescope. In Section 2 we discuss our observations of XO-2b. In Section 3 we detail our reduction procedure, which corrects for time varying systematics. In Section 4 we discuss the process used to analyze each of our light curves. Section 5 presents our coupled radiative transfer and photoionization model of the light curves. In Chapter 6 we explore the effects of Rayleigh scattering with the presence of optically thick clouds and discuss the implications of our derived Na abundance with respect to the structure of XO-2b's  ionosphere.

\begin{figure}
\centering
\hspace*{-0.25in}
\includegraphics[scale=0.55]{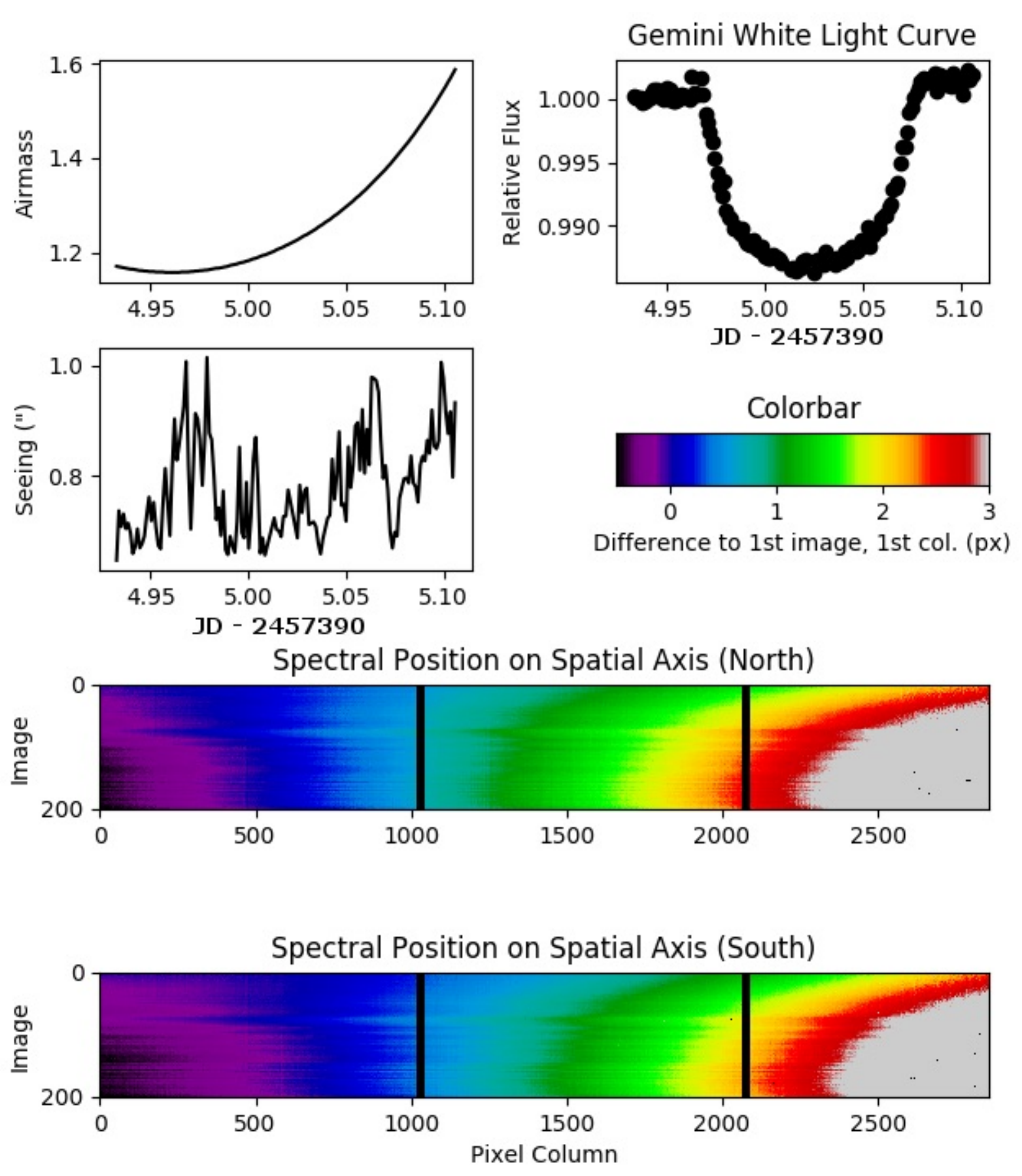}
\caption[GMOS Observation Stability]{ Various observing metrics including seeing, airmass and pointing stability are plotted for the duration of our Gemini observations. The white light curve (top right) is the wavelength averaged flux of XO-2N divided by XO-2S. Due to the nature of our large slits we find a direct correlation between changes in seeing and changes in resolution of our spectra. The bottom two plots show the change in spectral PSF centroid along the spatial axis (pixel columns) relative to the first image and first column of XO-2N and XO-2S, respectively. The colors are representative of translations along the spatial axis. Both XO-2N and XO-2S exhibit translations along the spatial axis that change with wavelength. The change in position is plotted as a function of image number in the y-axis. The spectra rotate over time relative to the first image, because the blue side of the detector (low pixel columns) exhibits a shift in the negative direction while the red side of the detector (high pixel columns) exhibit a shift in the opposite direction. The black vertical lines are caused by gaps in the 3 detector arrays for the GMOS instrument. }
\label{obs_stats_g}
\vspace{0.09in}
\end{figure}

\begin{figure}
\centering
\includegraphics[scale=0.6]{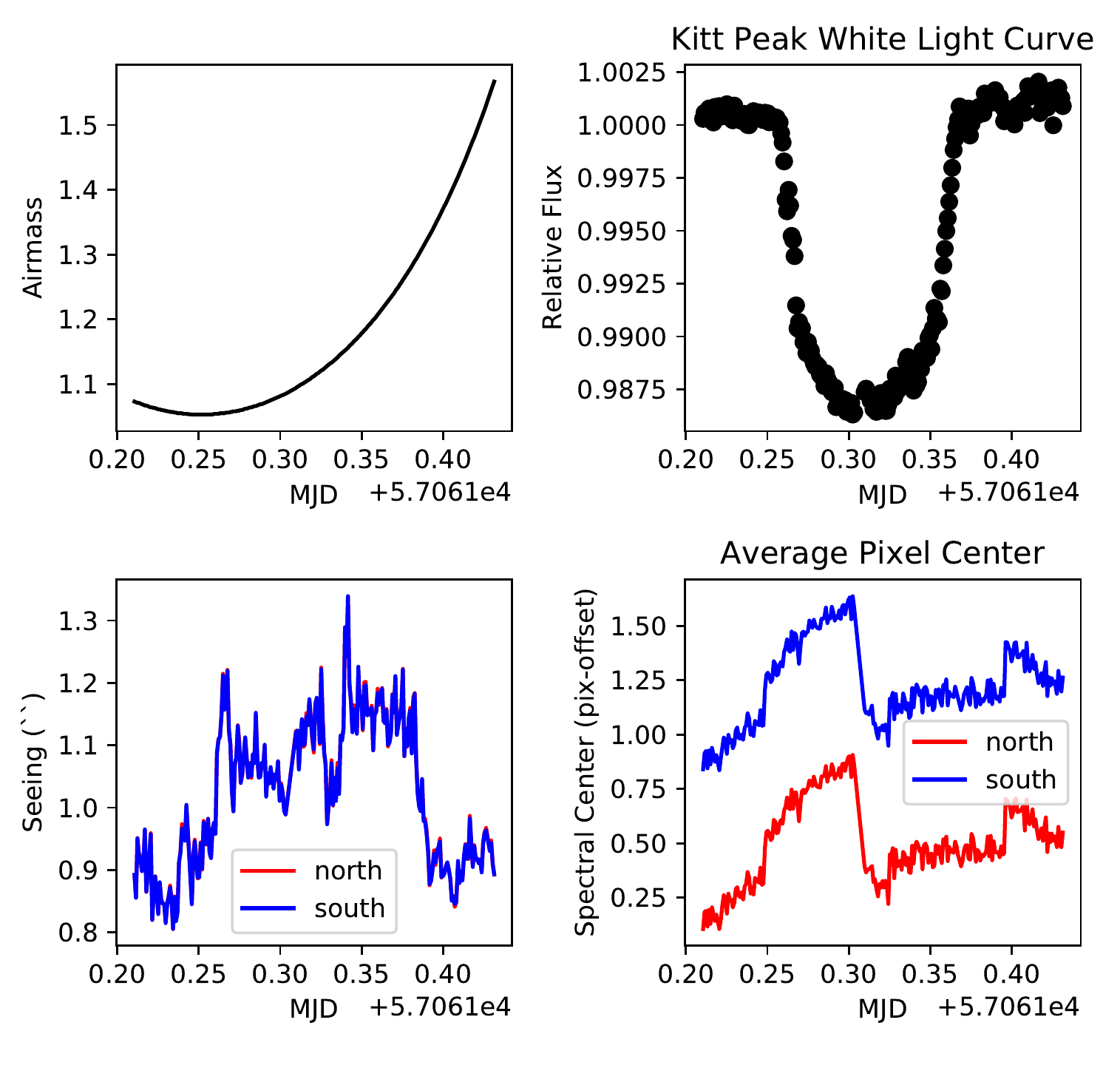}
\caption[KOSMOS Observation Stability]{ Various observing metrics including seeing, airmass and pointing stability are plotted for the duration of our Mayall observations. The white light curve (top right) is the wavelength averaged flux of XO-2N divided by XO-2S. The bottom right plot shows the average change in spectral PSF position along the spatial axis compared to the first image. We find a negligible 1 pixel difference in the PSF position across all wavelengths. The large jump in PSF position was the result of a loss in guiding that required a repositioning of the stars in the slit mask. }
\label{obs_stats_k}
\vspace{0.09in}
\end{figure}

\section{Observations}

\subsection{Gemini Data}
Spectroscopic measurements of XO-2b were recorded with GMOS-N on 2016 January 05 at the Gemini 8.1 m Observatory on Mauna Kea in Hawaii \citep{Allington2002}. The spectrograph is equipped with a e2v deep-depletion (DD) array of three CCDs which make a 6144$\times$4608 pixel sensor with a plate scale of 0.0727$\arcsec$ per pixel. The three CCDs are stitched together to create a large detector with a $\sim$20 pixel gap between each CCD. To reduce the read-out time to 12 seconds, we implemented 2$\times$2 binning and readout a subarray centered only on XO-2N and XO-2S. Observations were conducted with the B600 grating (R$\sim$1600) centered on a wavelength of 490~nm. Clocks were synchronized with a GPS every few seconds to ensure accurate timing. Seeing during the observations ranged from 0.64$\arcsec$ to 1.07$\arcsec$. The cadence of our observations was 102.5 seconds. Observations began 1 hour prior to the transit ingress and ended 1 hour after egress in order to characterize the out-of-transit baseline. 

A custom slit mask was created to perform simultaneous spectroscopy on XO-2N and its companion, XO-2S. To ensure that we captured the entire PSF of the two stars and to avoid any seeing-induced signal loss, we used 14$\arcsec$$\times$14$\arcsec$ wide slits. Small slits (2$\arcsec$$\times$5$\arcsec$) were placed on the outside of XO-2N and XO-2S and aligned with the respective larger slit to obtain a more accurate wavelength calibration. 

To gain further information on the wavelength alignment we used a CuAr lamp. A wavelength-pixel solution is created by discretizing a non-linear trend into 8 linear segments derived from interpolation between 8 known CuAr spectral features. Figure \ref{obs_stats_g} summarizes the seeing, dispersion and airmass during the observations. We find the dispersion of the spectrum rotates on the detector during our observations. The rotation, however, is negligible over 2000 pixels and no rectifying corrections are needed. However, the point spread function (PSF) in each column of the detector shifts relative to the PSF in the first column of the first image (Figure \ref{obs_stats_g}). The red channels (lower pixel columns) exhibit a -0.5 pixel shift in position relative to the first image, while the blue channels (larger pixel columns) exhibit a 3 pixel shift relative to the first image. XO-2S and XO-2N both exhibit a rotation but to varying extents. This rotation is negligible over our spectral extraction aperture because it produces at most a 0.12 pixel difference along the dispersion axis. 

\subsection{Mayall Data}
We obtained one transit observation of XO-2b with the 4 m Mayall Telescope on Kitt Peak in Arizona equipped with the multiobject visible-band spectrograph KOSMOS on 2015 February 07. The spectrograph is equipped with a 2k x 4k e2v Deep Depletion CCD with a binned plate scale of 0.292$\arcsec$/pixel \citep{Martini2014}. To reduce the read-out time to 19 seconds the CCD was binned 2x2. We used the Blue VPH grism, which has a wavelength range of 370--620~nm and R=2100 with peak transmission $\sim$0.4 near 500 nm. Clocks were synchronized with a GPS every few seconds to ensure accurate timing. Throughout the night the maximum shift in the centroid along the spatial axis on the detector of our target was less than 1 pixel due to adequate auto-guiding. Seeing in the observations ranged from 0.81$\arcsec$ to 1.34$\arcsec$ throughout the observations. We experienced a loss of guiding around 1.3 hours into the transit and had to reposition the stars on the slit. This repositioning created a minor mis-alignment of 0.5 pixels along the dispersion axis and roughly a 0.6 pixel shift along the spatial axis. Figure \ref{obs_stats_k} shows the airmass, seeing, and positioning of the spectra throughout our observations.

We created a specific slit mask for this observation and placed 5 large slits (10$\arcsec$ x 10$\arcsec$) around our region of interest: one around XO-2N, one around XO-2S, and three to probe the background flux between and outside of XO-2N and XO2-S. We did not include narrow slits as we did for our GMOS mask. Instead, we took two images of an HeNeAr lamp using a narrow long slit (1$\arcsec$ wide) and our science mask and then aligned the two for both XO-2N and XO-2S. To minimize the difference in spectral dispersion between our science image and arc lamp, we recorded a lamp image while the telescope is pointing at the XO-2 system and directly before the science images because it minimizes the affects of telescope flexure.

\begin{figure}
\includegraphics[scale=0.65]{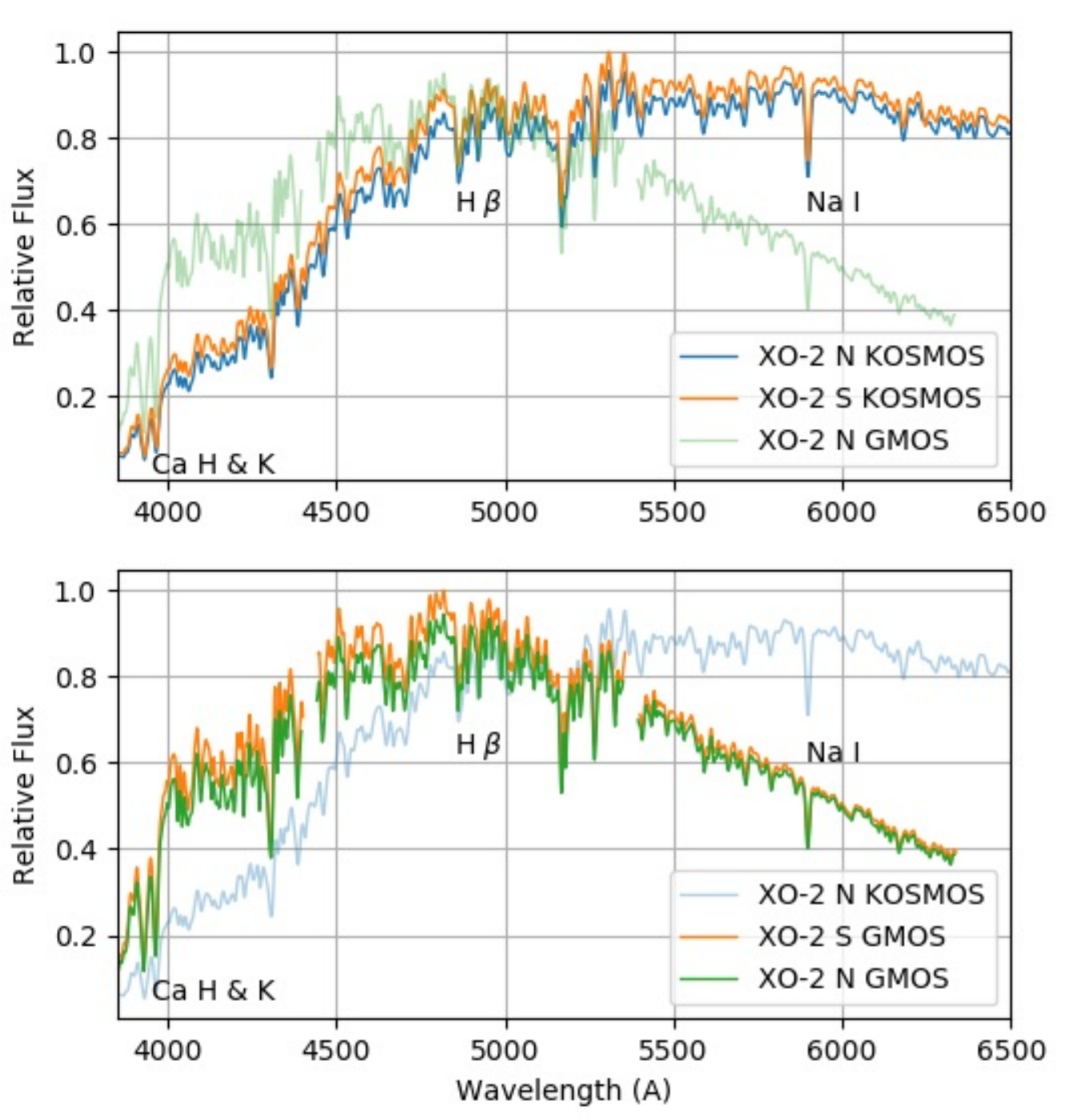}
\centering
\caption[stellar]{ \textcolor{black}{ Reduced spectra of the target and comparison star from each instrument. The spectra are normalized by the maximum flux value from XO-2 S. All spectra were extracted at similar airmass values. Since the pixels in the two detectors are different sizes, the spectra presented here are binned to the resolution of the observations (2 nm) for clarity in defining major spectral features. Prominent stellar features are labeled. The gaps in the wavelength coverage of the GMOS data are due to the physical gaps between individual CCDs in the detector. Neither spectra are flux calibrated but XO-2S has been wavelength calibrated such that the spectral features are aligned to those in XO-2N.} }
\label{PSF}
\vspace{0.09in}
\end{figure}

\section{Data Reduction}

Observations from GMOS and KOSMOS are reduced in the same manner unless otherwise stated. We extract the time-varying flux of each target (XO-2N and XO-2S) from every pixel column between 3800-6400 \AA~ using an aperture mask described below. All images are bias-subtracted and flat-fielded. We use 10 flat-field images with 2 second exposures to adequately characterize the detector illumination function. \textcolor{black}{There is an error associated with our flatfield correction since we did not correct for time dependent effects. However, this does not play a large role because the gradient of the flatfield spectrum is smaller than that of the stellar spectrum so shifting wavelengths has a smaller affect within each bin.} The flux is summed within each spectral channel using an aperture mask determined by the size of each slit. The aperture mask is 75 pixels along the spatial axis for GMOS and 30 pixels along the spatial axis for KOSMOS. The size of each aperture mask was determined with a flat field image because the flat field lamp fully illuminates each slit. We use a large aperture mask to acquire as much information about the flux from the star as possible. In order to correct for the time-varying sky background, we fit for the background flux in each image and wavelength using a PSF profile and fixed offset (Figure \ref{PSF}). For our GMOS data only, we use a pseudo-Voigt profile to model the PSF of the star and background simultaneously (Equation \ref{psvoigt}). We do not fit the PSF for KOSMOS because we have slits dedicated to measuring the background. When modeling the sky background, a Gaussian overestimates the background level and creates negative counts that are outside our read noise and a Lorentz profile underestimates the background (Figure \ref{PSF}). However, both a Gaussian and Lorentz profile fit the center of the PSF equally well but diverge at the wings, if the standard deviation is the same (See Figure \ref{PSF}). The pseudo-Voigt profile we use is

\begin{equation} \label{psvoigt}
(1-W) A e^{\frac{-(x-\mu)^{2}}{2\sigma}} + W \frac{A\sigma^{2}}{(x-\mu)^{2} + \sigma^{2}} + B
\end{equation}

\noindent
where A is the amplitude of the PSF, $\mu$ is the center, $\sigma$ is the standard deviation, $B$ is the background level, and $W$ is the weighted sum coefficient that mixes a Gaussian and Lorentz profile. The amplitude, standard deviation and mean parameters are shared between the Gaussian and Lorentz profile for consistency. We subtract the background, $B$, from each pixel after modeling the PSF using a constrained least-squares minimization of the percent error. A minimization of the percent error provides a better fit to the wings than the chi-squared. If the pixel values span orders of magnitude, the chi-squared will focus on minimizing large deviations in the PSF peak and ignore smaller variations in the wings. 

\begin{figure}
\includegraphics[scale=0.50]{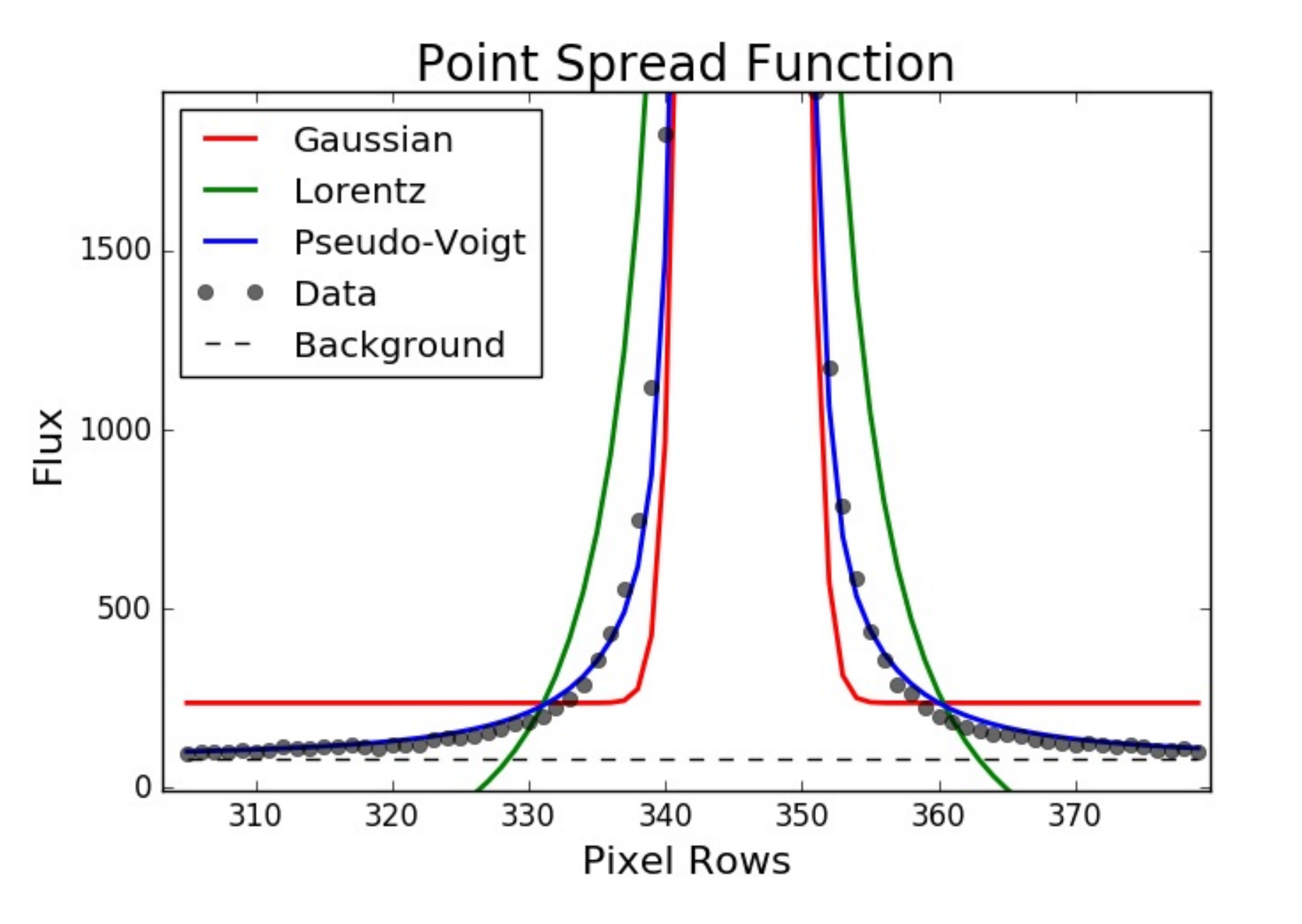}
\centering
\caption[PSF]{ The point spread function (PSF) of XO-2S from the first column of the first image is shown from our GMOS data. To estimate the background flux of the data, we simultaneously model the PSF with a vertical shift added to a Gaussian, Lorentz or Pseudo-Voigt profile. Each model agrees with the center of the PSF but differs greatly at the wings. We find a weight (W) of 0.272 is needed to accurately model the PSF with the Pseudo-Voigt profile (see Equation \ref{psvoigt}). The Pseudo-Voigt profile yielded a background flux (B) of $\sim$80 counts/pixel where as the Gaussian yielded 235 and Lorentz -338 e$^{-}$/pixel. We use the background estimate from the Pseudo-Voigt profile for our background subtraction. Each model is binned to the resolution of the data. }
\label{PSF}
\vspace{0.09in}
\end{figure}

\subsection{Wavelength Calibration}

\begin{figure*}[ht]
\centering
\includegraphics[scale=0.5]{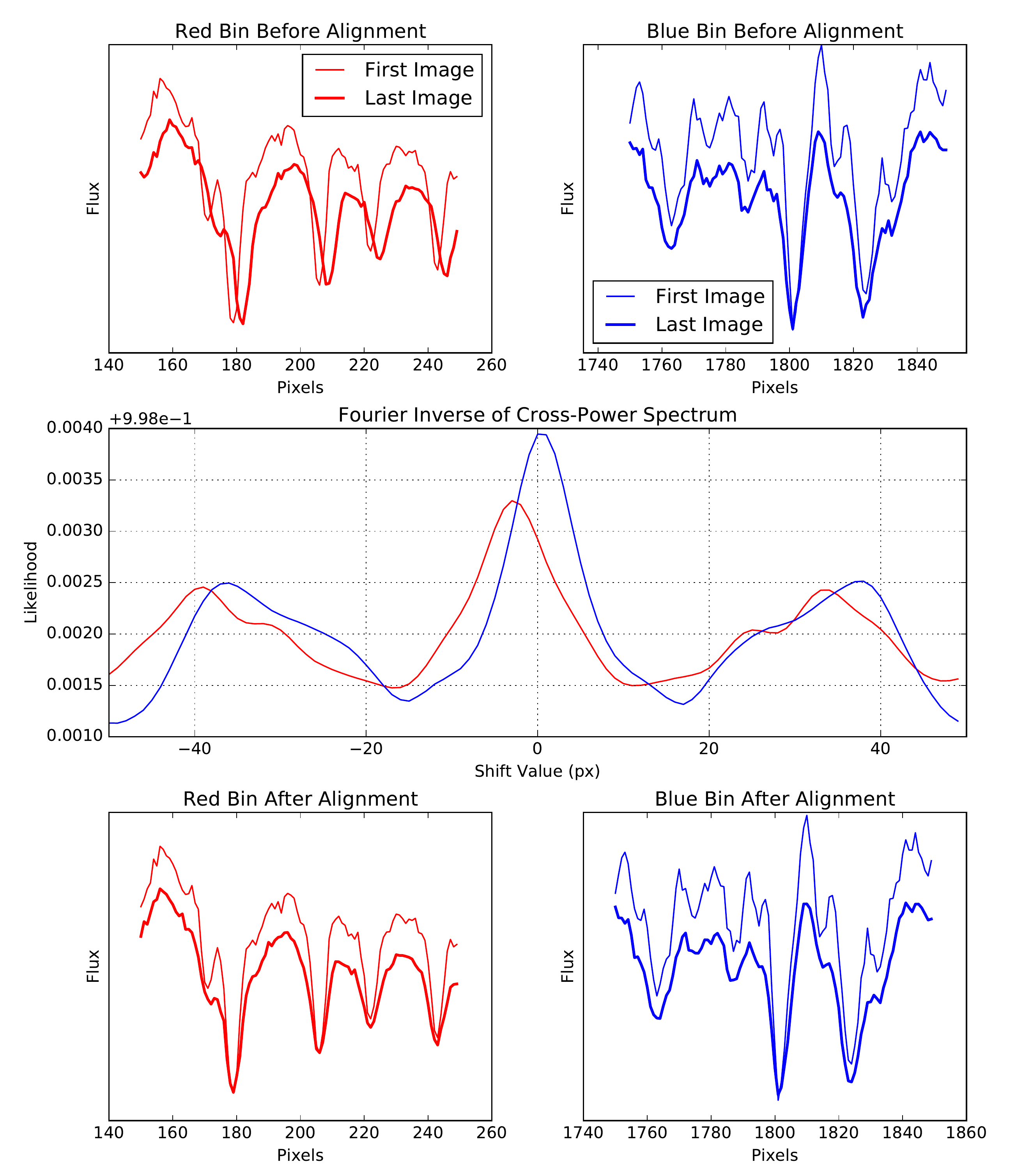}
\caption[GMOS Wavelength Calibration]{
\textcolor{black}{
In order to detect a need for a time-dependent wavelength calibration we cross-correlate the first and last image on two sides of our detector. \textbf{Top}, spectra of XO-2 N from GMOS on two different sides of the detector. Visually, you can see an offset between the first and last image on the red side of the detector but the shift is more subtle on the blue side. If we were to apply the same wavelength calibration to the first and last image we would introduce systematic errors because the spectral features are not aligned. \textbf{Middle}, a cross-power spectrum from the cross-correlation algorithm which shows each side of the detector requiring a different offset for alignment, as indicated by the maximum of each curve. \textbf{Bottom}, aligned spectra using the offset derived from the POC (see Equation \ref{poc}). Since the shift is wavelength dependent we can not use a single cross-correlation on our data to align it in time, each wavelength bin must be treated separately. }
}
\label{obs_cc}
\vspace{0.09in}
\end{figure*}

\begin{figure}[h]
\centering
\includegraphics[scale=0.55]{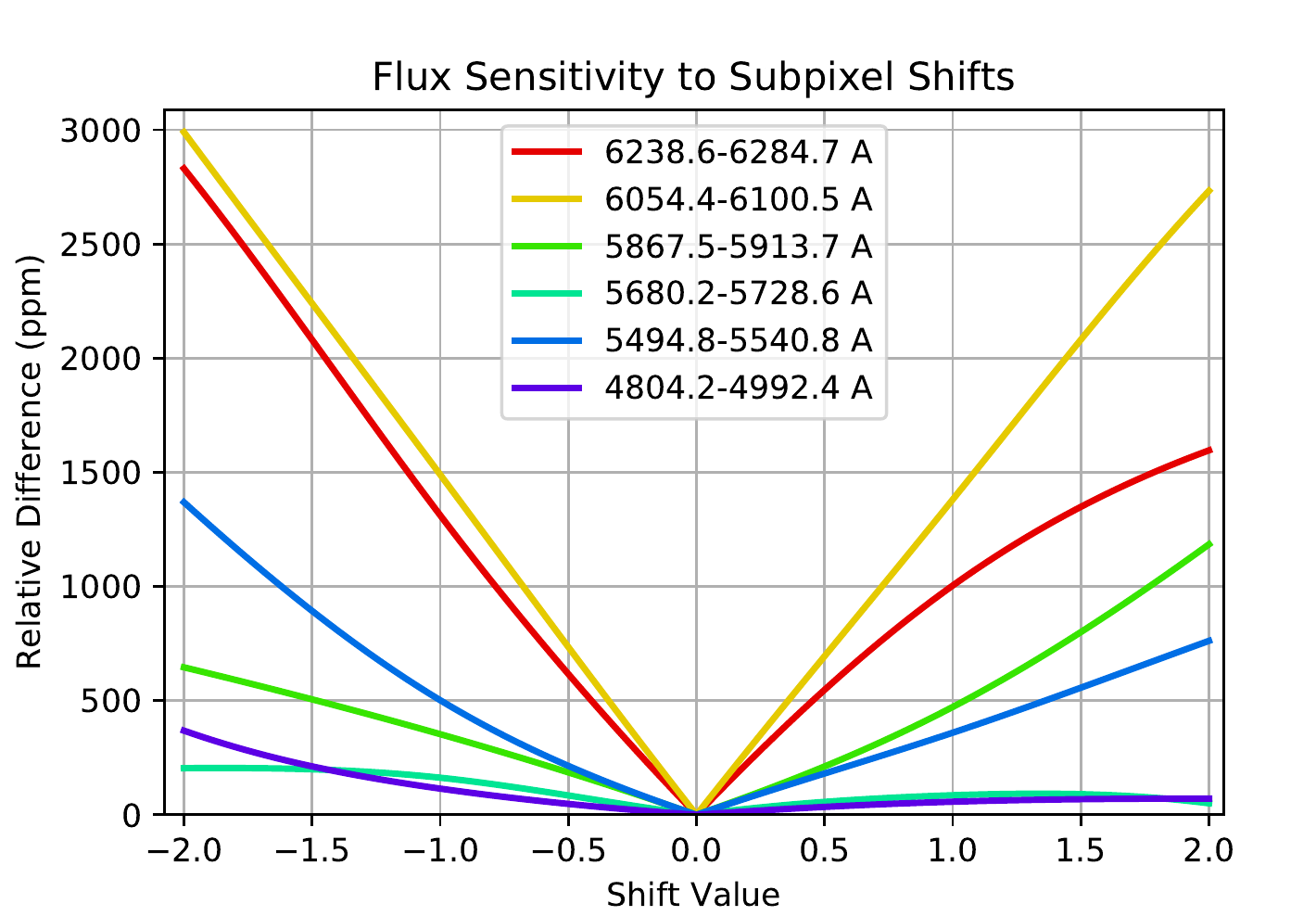}
\caption[Shift Sensitivity]{ The relative change in flux of XO-2N for misalignments within a wavelength bin on the order of 2 pixels using the Gemini data. The change in flux is computed as the ratio of bin integrated fluxes between the original (no offset) bin and a bin that is slightly shifted from the original position. Bins with larger flux gradients at the edges exhibit larger differences under misalignments. The cross correlation of wavelength bins should be done at a resolution larger than the data to achieve subpixel precision. }
\label{shift_sens}
\vspace{0.09in}
\end{figure}

	The absorption spectrum of the exoplanet was extracted using the common technique of dividing the light of the host star by one or more reference stars. Here we use only the reference star XO-2S, because it is exceptionally similar in brightness and stellar type to the host star, to divide out terrestrial atmospheric effects, the stellar spectrum, as well as the systematic errors, e.g. telescope jitter. \textcolor{black}{ XO-2 N is a G9V star with a magnitude of 11.138 V and a 2MASS identifier of J07480647+5013328. Located only 31.1" away in the sky is XO-2 S, a G9V star with a magnitude of 11.086 V and a 2MASS identifier of J07480748+5013032} \citep{Damasso2015}. Despite the similarities, we find that the division of the XO-2N$/$XO-2S spectra is highly sensitive to time-varying wavelength shifts, which change with wavelength (Figures \ref{obs_cc} and \ref{shift_sens}). 
    
	Temporal variations in the spectroscopic dispersion correlate with changes in seeing, airmass variations, and, telescope flexure. Our dataset experiences changes in resolution due to changes in seeing, and translations and stretching along the dispersion axis. Since we use large slits to capture all the flux from the stars, the dispersion of the data is dictated by the seeing. As the seeing increases, the resolution of the spectral features decreases. In addition, the non-linear translations along the dispersion axis cause the spectral features to move farther apart from one another over time. The consequential spectral misalignment between the host and reference stars significantly affects the resultant derived spectrum, by both masking existent features and creating false signatures. While aligning data has been realized in past literature (e.g. \citealt{Waldmann2012}; \citealt{Pearson2016}; \citealt{Huitson2017};) methods for correcting and tracking such systematics have not been discussed in depth.
    
In order to align the host star’s spectrum to that of the reference star, we use a phase correlation algorithm that achieves subpixel precision for aligning 1-dimensional signals. Phase correlation (\citealt{Kuglin1975}) is a common technique for image alignment with broad applications in image stitching and computer vision \citep{Castro1987}. The method relies on finding the maximum of the phase-only correlation function (POC) which is defined as the inverse Fourier transform of the normalized cross-power spectrum between two signals. The coordinate of the maximum in the POC, $\mathcal{P}$, corresponds to the translation between two signals:

\begin{equation} \label{poc}
\mathcal{P} = \mathcal{F}^{-1} \left( \frac{ \mathcal{F}(x_{1}) \mathcal{F}^{*}(x_{2}) }{|\mathcal{F}(x_{1}) \mathcal{F}^{*}(x_{2})|} \right)
\end{equation}

\noindent We cross-correlate only the data within each wavelength bin ($\sim$1.6 nm). represented as $x_{1}$ and $x_{2}$, where $\mathcal{F}$ is the Fourier transform, $\mathcal{F}^*$ is the complex conjugate of the Fourier transform and $\mathcal{F}^{-1}$ is the inverse Fourier transform.

\begin{figure}[h]
\centering
\includegraphics[scale=0.5]{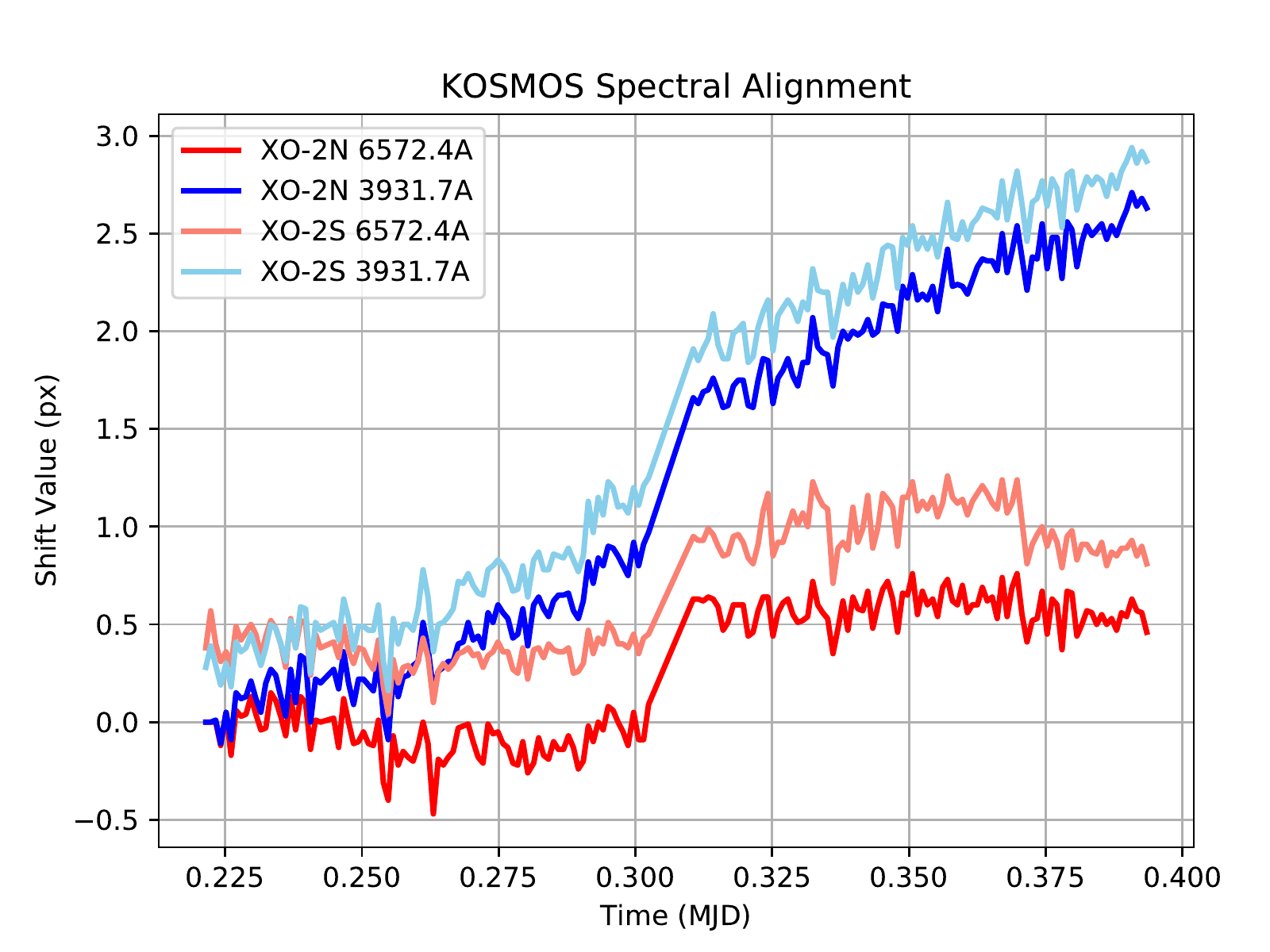}
\includegraphics[scale=0.5]{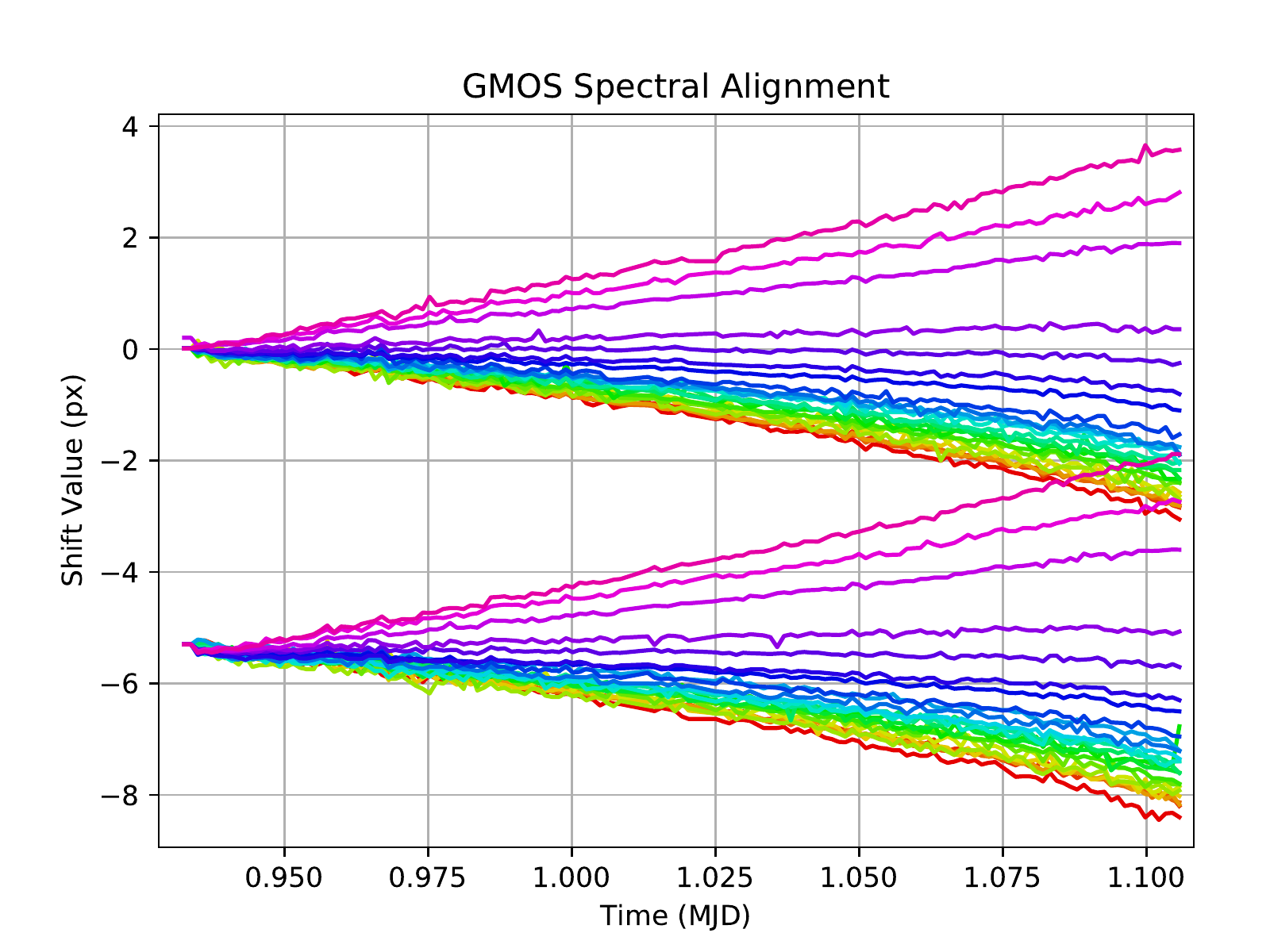}
\includegraphics[scale=0.5]{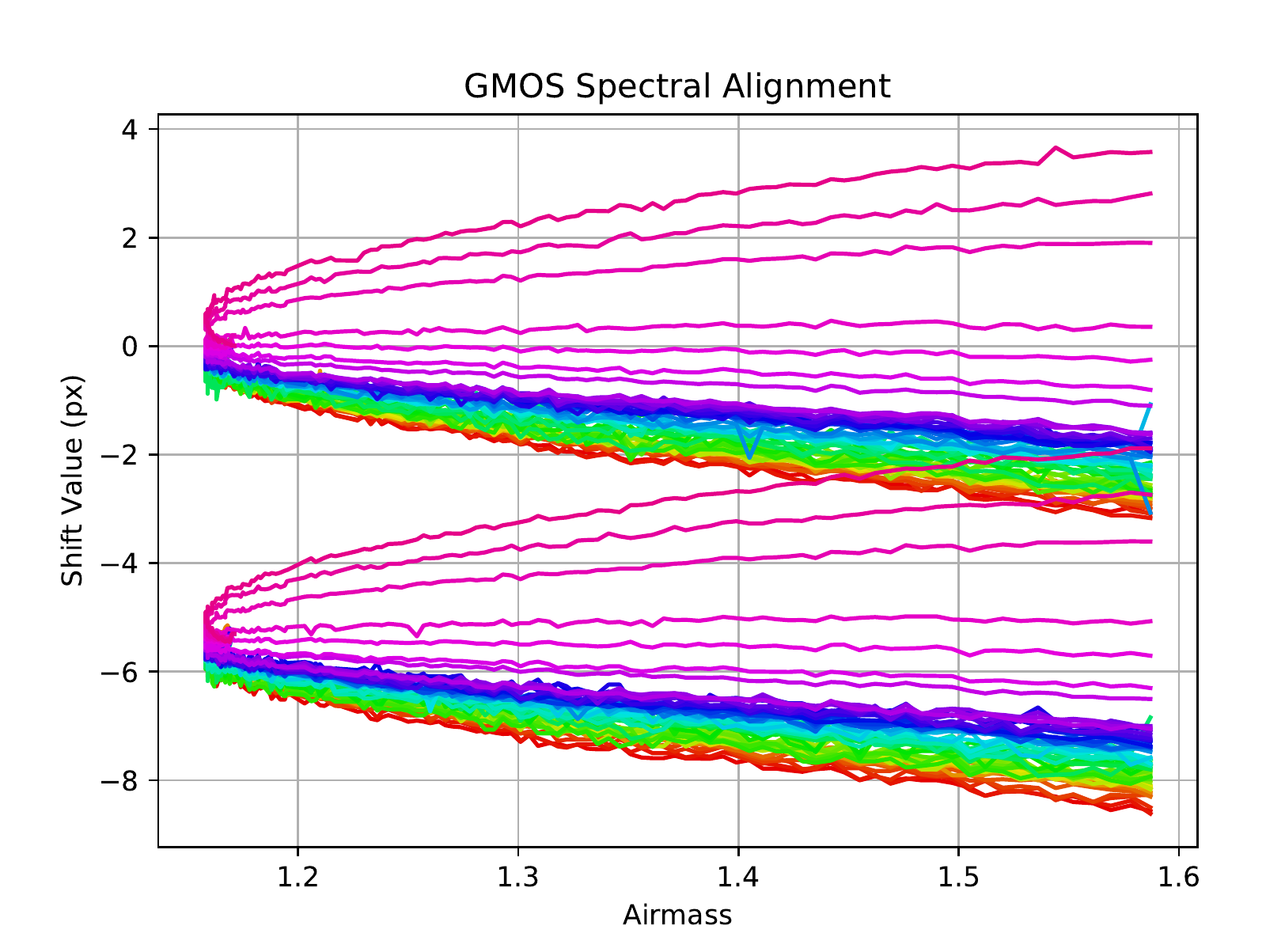}
\caption[Cross Correlation Shifts]{ The pixel shift required to align each wavelength bin to the template spectrum (the first spectrum of XO-2N). Both GMOS and KOSMOS spectra experience a shift over time. We hypothesize that atmospheric differential refraction causes most of the changes over time (Figure 1 on url\footnote{http://www.gemini.edu/sciops/instruments/gmos/itc-sensitivity-and-overheads/atmospheric-differential-refraction}). In the bottom subplot the color represents roughly the respective wavelength of the bin (as indicated in the appendix for the wavelength legend). The alignment data for XO-2S is shown as the lower set of lines in the bottom subplot. The large displacement or bias shift between XO-2N and XO-2S is due to the misalignment of the slits on the mask. KOSMOS did not experience as large of an alignment correction as did GMOS, therefore for clarity only the red-most and blue-most corrections are plotted and every other bin falls in between. }
\label{cc_shift}
\vspace{0.09in}
\end{figure}

A caveat to cross-correlating data at the native resolution is that the location of the maximum can only be obtained with integer precision. We mitigate this limitation by linearly interpolating the spectra onto a higher resolution grid (10$\times$ more points) to achieve subpixel precision with the cross-correlation. After cross-correlation and an optimal offset between the two signals is found, the target signal (at native resolution) is shifted with a linear interpolation to align with the template signal. Before we create each light curve, spectra of XO-2N are aligned with the first spectrum of XO-2N and spectra of XO-2S are aligned with the first spectrum of XO-2N. This alignment process is performed separately for each wavelength bin, where only the data in each bin is phase correlated. Figure \ref{cc_shift} shows the pixel translation between each wavelength bin for every image. The Python code for our signal alignment routine is provided online on GitHub\footnote{\url{https://github.com/pearsonkyle/Signal-Alignment}}.

We validate our alignment method against an alternate method that minimizes the chi-squared between wavelength bins. The data in each bin are first normalized (mean=1) so that we can compare their shapes. Then, a minimization of the chi-squared derives the optimal shift necessary to align the spectral features. This method achieves arbitrary subpixel precision without needing to interpolate the data onto a higher resolution grid. Our bin sizes, particularly around the Na feature, do not contain enough information to be aligned with 10 pixels. For instance, both alignment algorithms can become biased if the shift is comparable to the distance between spectral features and the bin width is comparable to the width of a spectral feature. These two conditions create a degenerate solution set because we could align two spectral features at different wavelengths since their shapes are similar. The POC method also succumbs to this fault because it inherently assumes periodic signals. We mitigate this degeneracy by increasing the amount of information in each bin by including extra data beyond the edges. We extend the bounds of our bins by an extra 20 pixels (10 pixels on top, 10 on bottom) with data from the spectrum only during the alignment stage.

\begin{figure}[h]
\centering
\hspace*{-0.25in}
\includegraphics[scale=0.4]{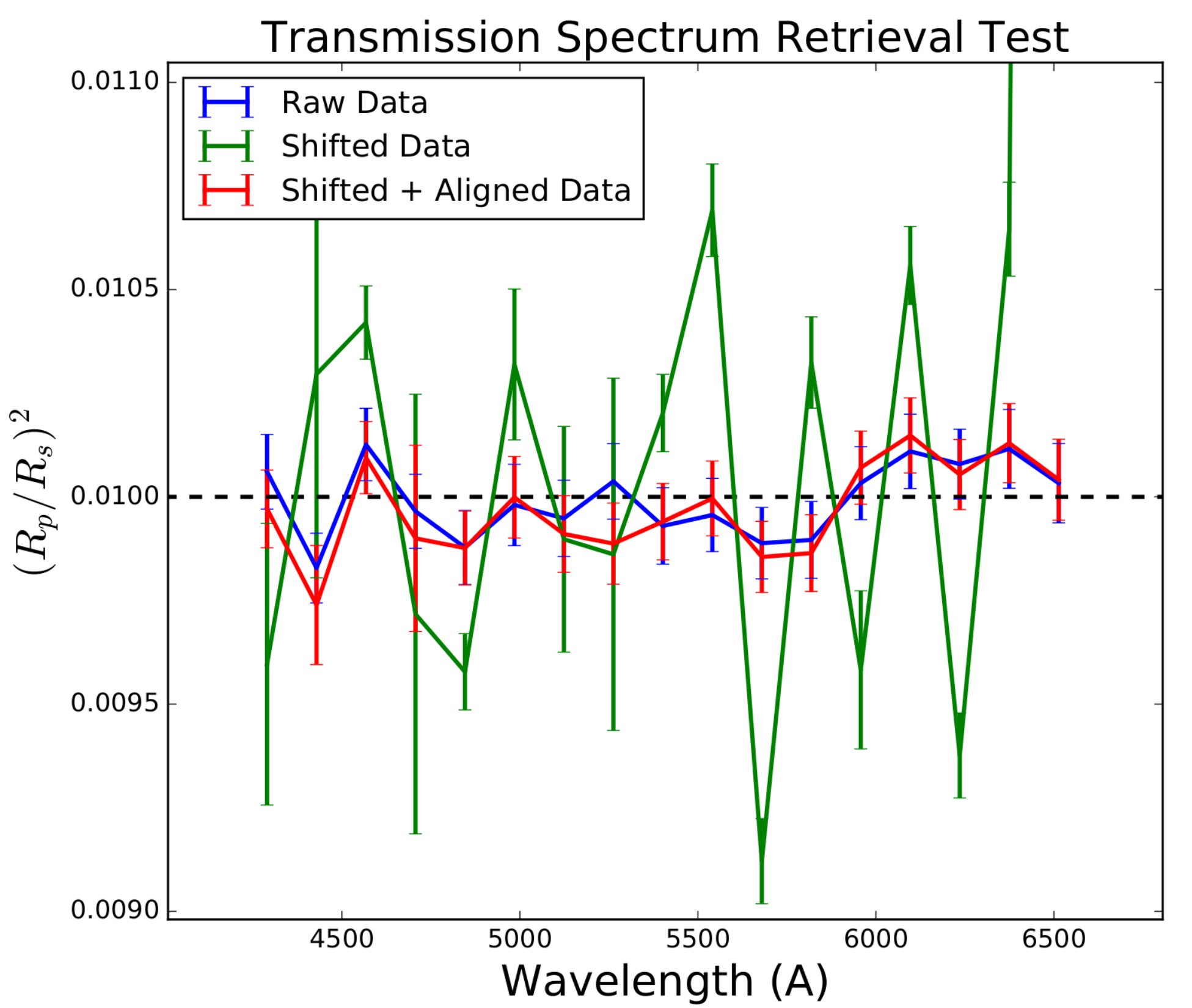}
\caption[Phase Correlation Null-Hypothesis Test]{ We verify our phase correlation algorithm with a retrieval test by simulating a dataset with a known transit depth and introduce a shifting systematic. We use a KOSMOS spectrum of XO-2N and inject a constant transit depth ($R_{p}/R_{s}$ = 0.1) into every wavelength and add Gaussian noise on the order of 500 ppm. The unshifted, noised up data is referred to as ``raw''. The raw data is linearly shifted up to 4 pixels between the first and last image and referred to as ``shifted''. The shifted data is then aligned using a phase correlation and compared to the original raw data. We find the phase correlation algorithm can successfully correct for a shifting systematic while the unaligned data shows up to $\sim$5\% change in transit depth.  }
\label{null-hypo}
\vspace{0.09in}
\end{figure}

We further validate the alignment algorithm with a retrieval test on simulated data. A transit signal is injected into a simulated data set along with a wavelength shift, similar to that found in our observations. We use the first stellar spectrum from our KOSMOS dataset as the model spectrum and create 150 copies (similar to our observations) to represent all of the images in our fake data set. A flat transmission spectrum was then multiplied into the data set using a $R_{p}/R_{s}$=0.1 and $u_{1}$=0.3 (linear limb darkening coefficient). To simulate the scatter of real observations, Gaussian noise on the order of $\sim$500ppm is introduced into each pixel column. Afterwards, we introduce a shifting systematic where the shift is linear and the difference between the first and last image is 4 pixels. We then derive the transmission spectrum for the shifted data to assess the extent to which the transit depth can change due to inadequately calibrating the data. Lastly, we align the data using our POC algorithm and derive the final transmission spectrum. Figure \ref{null-hypo} shows the results of our test. We find that our alignment algorithm calibrates the shifted data set to less than a 0.1$\sigma$ difference from our original solution. Therefore, we assume it can reliably calibrate our real observations.

As a result of the differing slit positions and optical paths that the light takes through the detector XO-2N requires a different wavelength correction than does XO-2S. Each star exhibits a wavelength dependence on the shift lengths which create a minor stretching effect over time. Thus, performing a cross-correlation on the whole spectrum over time will inaccurately align the data. Instead, each wavelength bin or region should be handled separately to avoid introducing systematic offsets into the data. In the GMOS dataset, the pixel difference between the first and last image for the red-most wavelength bin of XO-2N is -3.2 pixels. The blue-most wavelength bin of XO-2N experiences a pixel difference of +0.85 pixels. Figure \ref{cc_shift} shows the shifts required to align each wavelength bin over time.  Without aligning the data, we find that the normalized flux within each wavelength bin can change up to 2000 ppm or more depending on the rate of change of flux at the bin edges (i.e. is it flat or steep at the edge) (see Figure \ref{shift_sens}).

\textcolor{black}{KOSMOS experiences a wavelength dependent drift in a manner similar to Gemini/GMOS (i.e. bluer wavelengths require a larger correction). Since the physical pixel sizes and optics are different between the two instruments this leads to different corrections even considering the observations spanned similar airmass values.} The red-most bin of the KOSMOS data experiences a pixel difference of +0.6 pixels between the first and last XO-2N spectrum. For the blue-most bin of the KOSMOS data we find a pixel difference of +2.6 pixels between the first and last XO-2N spectrum. The largest shift required to align the data occurred when we lost guiding and had to reposition the telescope as shown in Figure \ref{cc_shift}.

\section{Analysis}

\subsection{Light Curve Model}

The planetary and orbital parameters of XO-2 b are derived by fitting a light curve model to the observations. Specifically, we assume prior measurements of the orbital period, inclination, scaled semi-major axis, and eccentricity listed in Table \ref{tab:fixedpars}. Our analysis then derives the planet-to-star radius ratios ($R_{p}/R_{s}$), linear limb darkening coefficients ($u_{1}$) and airmass correction factors ($u_{0}, a_{1}$). Additionally, the time of mid-transit ($T_{mid}$), left as a fixed parameter for each wavelength bin, is acquired by leaving it as a free parameter when modeling the white light curve. We use the analytic expressions of \cite{Mandel2002} to generate our model transits, modified to account for ground-based observations with atmospheric extinction. The following function below is used to maximize the likelihood of the transit model and airmass signal simultaneously:

\begin{equation} \label{expam}
F_{obs} = a_{0} e^{a_{1} \beta }  F_{transit}
\end{equation}

\noindent
Here $F_{obs}$ is the flux recorded on the detector, $F_{transit}$ is the actual astrophysical signal (i.e. the transit light curve, given by a Mandel and Agol model light curve), $a_{i}$ are airmass correction coefficients and $\beta$ is the airmass value. For each wavelength bin, the flux of XO-2N is divided by the flux of XO-2S to remove shared systematic errors, the largest of which is airmass. The prior step removes most of the systematics between the two stars but leaves a residual curvature in the data due to the fact that both stars do not share exactly the same systematics (e.g., atmospheric optical path and pixel-sensitivity of the instrument). We model the residual curvature in each wavelength bin using an exponential extinction function. The coefficients for the exponential airmass function are left as free parameters along with the transit parameters of interest, e.g., $R_{p}$/$R_{s}$.

As the planet transits in front of the host star, brightness contrasts between the stellar limb and center modulate the shape of the transit. Uncertainties on our flux measurements range from $\sim$500 -- 1500 ppm  and are not precise enough to resolve the difference between linear and quadratic laws. As a quadratic profile would introduce additional degeneracies when fitting our current data, we correct for limb darkening (LD) effects in our transit model with a linear profile \citep{Schwarzschild1906}. To enable a robust retrieval of the linear LD coefficient we keep it as a free parameter during our analysis and constrain it to within the uncertainties derived from the stellar parameters (i.e. T$_{eff}$, $\log{g}$, $[Fe/H]$).

We use a nested sampling algorithm and constrain the simulation by using an initial fit to the data from a non-linear least-squares (LS) constrained minimization (\citealt{Branch1999}; \citealt{scipy}). A constrained minimization is used to allow for physically sensible values, i.e., the radius of the planet will not be larger than the radius of the star and the mid transit time is within our observation window. 

The likelihood function, used to assess how well a model fits the data, is calculated with the $\chi^{2}$ of the model modified by a loss function. We ignore data points greater than 3-sigma away from the model because it reduces the influence of outliers on the solution. Since the uncertainties on the data points are $\sim$5--10 times smaller than the shifting sensitivities we minimize the influence of misalignments by ignoring any data points greater than 3-sigma away from the model fit as well.

\begin{table*}
  \caption{Fixed Model Parameters}
  \label{tab:fixedpars}
  \begin{center}
    \leavevmode
    \begin{tabular}{lll } \hline \hline
  $Parameter$  &Value  & Reference        \\ \hline

\textbf{XO-2b's Orbital Parameters} & & \\
 Period (days)							&   2.61586178    &  \citet{Sing2011}   \\
 Inclination ($\degr$)  					&   88.01             & \citet{Crouzet2012}  \\
 a/R$_{s}$   							&   7.986	           & \citet{Crouzet2012} \\
 Eccentricity						&   0	& \citet{Crouzet2012}		   \\
 T$_{mid}$ - GMOS (JD) & 2457395.02261 & White Light curve Fit \\
 T$_{mid}$ - KOSMOS (JD)  & 2457061.81231 & White Light curve Fit \\
& & \\

 \textbf{Host Star XO-2N Parameters} & & \\
T$_{eff}$				&	5440 $\pm$ 69 K			& \citet{Teske2015}\\
$\log{g}$ (cgs)			&	4.35 $\pm$ 0.19		    & \citet{Teske2015}\\
$[Fe/H]$				&	0.45 $\pm$ 0.06			& \citet{Teske2015}\\

    \end{tabular}
  \end{center}
\end{table*}

 In order to find a global fit solution, we employ the use of the multimodal nested sampling algorithm called MultiNest (\citealt{Skilling2006}; \citealt{Feroz2008}; \citealt{Feroz2009}). MultiNest is a Bayesian inference tool that uses the Monte Carlo strategy of nested sampling to calculate the Bayesian evidence alongside enabling posterior inference, thereby allowing simultaneous parameter estimation and model selection. A nested sampling algorithm is efficient at probing parameter spaces which could potentially contain multiple modes and pronounced degeneracies in high dimensions; a regime in which the convergence for traditional Markov Chain Monte Carlo (MCMC) techniques becomes incredibly slow (\citealt{Skilling2004}; \citealt{Feroz2008}). Moreover, MCMC methods often require careful tuning of the prior distribution to sample efficiently, and testing for convergence can be problematic.

The nested sampling algorithm works by first drawing N samples from the full prior which is simply a uniform distribution over the whole prior range ($\pm$10$\sigma$ away from the initial fit with the LS technique). The samples are then sorted in terms of their likelihood and the smallest likelihood ($L_{0}$) is removed from the ``live set". The removed point is then replaced by a point drawn from the prior subject to the constraint that the point has a likelihood larger than the previous minimum, $L_{0}$. At each subsequent iteration i, the discarding of the lowest-likelihood point $L_{0}$ in the live set, the drawing of a replacement with $L_{0}$ $\le$ $L_{i}$ and the reduction in the corresponding prior volume are repeated, until the entire prior volume has been traversed. The algorithm thus travels through nested shells of likelihood as the prior volume is reduced. The stopping criterion for the algorithm is when the remaining prior volume and maximum-likelihood value would no longer change the final Bayesian evidence estimate to within some user specified tolerance. Once the evidence is found the posterior inferences can easily be generated using the full sequence of discarded points from the nested sampling process. An in depth description of how MultiNest searches the parameter space and constructs its elliptic bounds to reduce the prior volume can be found in Section 5 of \cite{Feroz2009}.

Optimization of the hyperparameters for nested sampling enable us to find the global solution in an efficient manner while producing a numerical uncertainty from sampling the posterior distribution. We use 500 live points with an evidence tolerance of 0.1 and a sampling efficiency of 10\% to ensure enough points in our prior space are sampled for convergence and to reveal any multimodal posterior distributions. Our prior range is defined as a hypercube with 4 dimensions (i.e. $R_{p}/R_{*}$, $u_{1}$, $a_{0}$, $a_{1}$) corresponding to our free parameters. The hypercube is centered on the values returned from the least-squares fit with bounds that extend 10$\sigma$ from the center of the cube. The $\sigma$ values used to define the prior hypercube size are determined from the preliminary least-squares fit. The results are shown in Table \ref{tab:finalpars} where u$_{1}$ is the linear limb darkening coefficient, $\sigma_{res}$ is the residual standard deviation and $a$ are the airmass coefficients shown in equation \ref{expam}.

\begin{figure*}[h]
\centering
\hspace*{-.25in}
\includegraphics[scale=0.80]{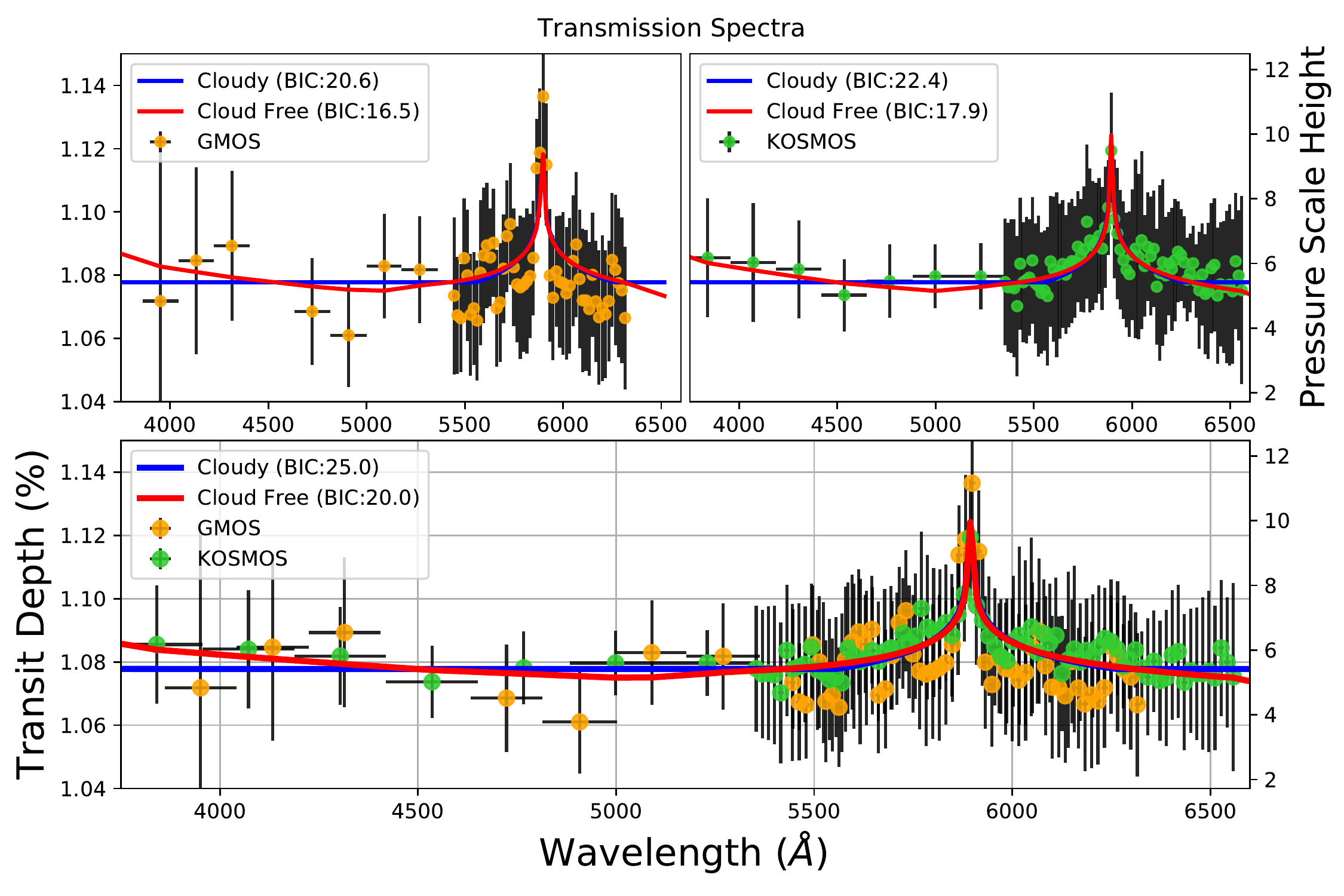}
\caption[XO-2b Transmission Spectrum]{ The transmission spectrum of XO-2b from GMOS and KOSMOS is shown along with our best fit atmospheric models. Each atmospheric model assumes the temperature structure and Na mixing ratio given in Figure \ref{atmo_model}. The atmospheric models are binned to the resolution of the respective observational data. We measure an absorption depth of 459$\pm$120 ppm ($\sim$ 4 scale heights) associated with the Na feature at 589 nm. The ``Cloudy'' model has an optically thick cloud at $\sim$ 30 mbar, a larger Na abundance and smaller 10-bar radius than the ``Cloud Free'' model. It is important to note that while the chi-squared of the ``Cloudy'' and ``Cloud Free'' models are the same, the BIC is different due to the dependence on the number of model parameters. The BIC values are computed using the total wavelength region (despite only showing a subset in one plot). 
}
\label{transmission_comparison}
\end{figure*}

\subsection{Radiative Transfer and Photoionization Model}

The interpretation of transmission measurements from close-in exoplanets requires a coupled model of photoionization and radiative transfer. The atmospheric transmission model is computed with 80 vertical layers from a uniform log pressure grid between 10 and 1e-7 bar. We include opacity sources due only to neutral Na and Rayleigh scattering of H$_{2}$. Due to the relative ease of ionizing Na, we compute the ion density in each atmospheric layer by assuming equilibrium between ionization (i.e. thermal- and photo-) and recombination (i.e. radiative and 3-body) in this equation:

\begin{equation} \label{photoionization}
\begin{split}
\frac{d}{dt}n_{Na^+} & =  \\
(Photoionization) &~ n_{Na} \sum_{\lambda} \sigma_{\lambda} F_{\lambda} e^{-\tau_{\lambda}} \\
(Thermal-Ionization)  &~ + <\sigma v>n_{e} n_{Na} \\
(Radiative~Recomb.)  &~ - \alpha_{r} n_{e} n_{Na^+} \\
(3~body~Recomb.) &~ - \frac{k_0 k_1 n_{H_2}}{k_0n_{H_2} + k_1} n_{e} n_{Na^+} 
\end{split}
\end{equation}.


Here $n_{Na^+}$ is the density of Na ions, $n_{e}$ is the electron density, $n_{Na}$ is the neutral Na density, $\alpha_{r}$ is the radiative recombination rate coefficient \citep{Verner1996a}, $\sigma_{\lambda}$ is the photoionization cross section \citep{Verner1996b}, $F_{\lambda}$ is the photon flux at XO-2b,  $\tau_{\lambda}$ is the optical depth from the top of the atmosphere down to a respective layer and the $\lambda$ subscript represents each quantity that is dependent on wavelength. Additionally, $<\sigma v>$ is the rate coefficient for thermal ionization \citep{Voronov1997}. The last term in equation \ref{photoionization} represents three body recombination \citep{Baulch2005} with rate coefficients $k_0$=3.43e-14*$T^{-3.77}$ and $k_1$=1e-7 \citep{Su2001} and, $n_{H_{2}}$ is the background density. We use a photon flux from the Sun (G2V) but scaled to the orbital distance of XO-2b and only integrate between 1--241.2 nm, where the upper bound represents the ionization limit for Na. We assume that the only source of electrons derive from the Na ions such that $n_{e}$ = $n_{Na^+}$.

\begin{figure*}[ht]
\centering
\includegraphics[scale=0.80]{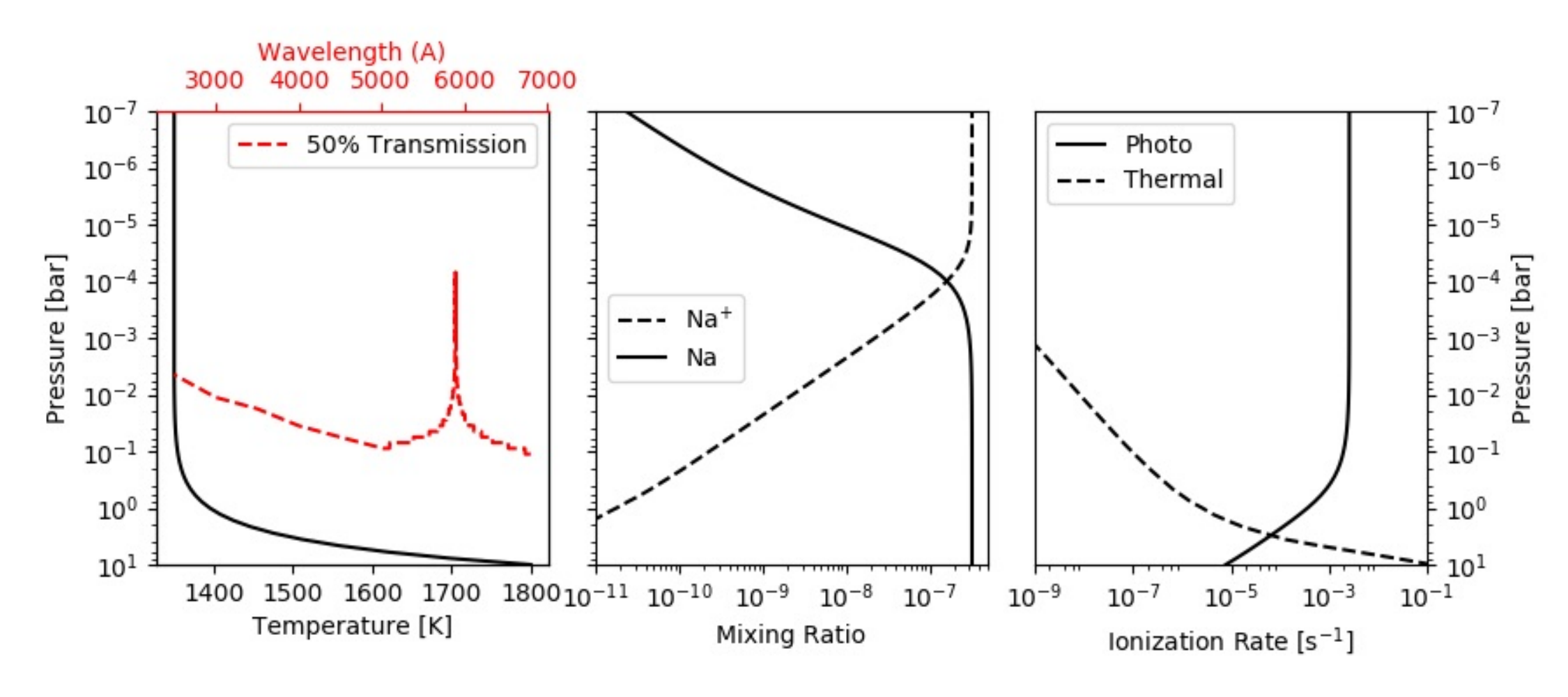}
\caption[XO-2b Photoionization]{ The atmospheric structure of XO-2b, derived from our grid search for a cloud free atmosphere. The Na mixing ratio is derived by balancing ionization and recombination (Equation \ref{photoionization}). The red dotted line indicates the level of 50$\%$ transmission. To generate this model we use an initial Na mixing ratio of 4.08e-7 and a temperature profile transition at 1e-1 bar above which the atmosphere is isothermal at 1350 K and below which decreases log-linearly to 1800 K at 10 bar. }
\label{atmo_model}
\end{figure*}

\begin{figure*}[ht]
\centering
\includegraphics[scale=0.80]{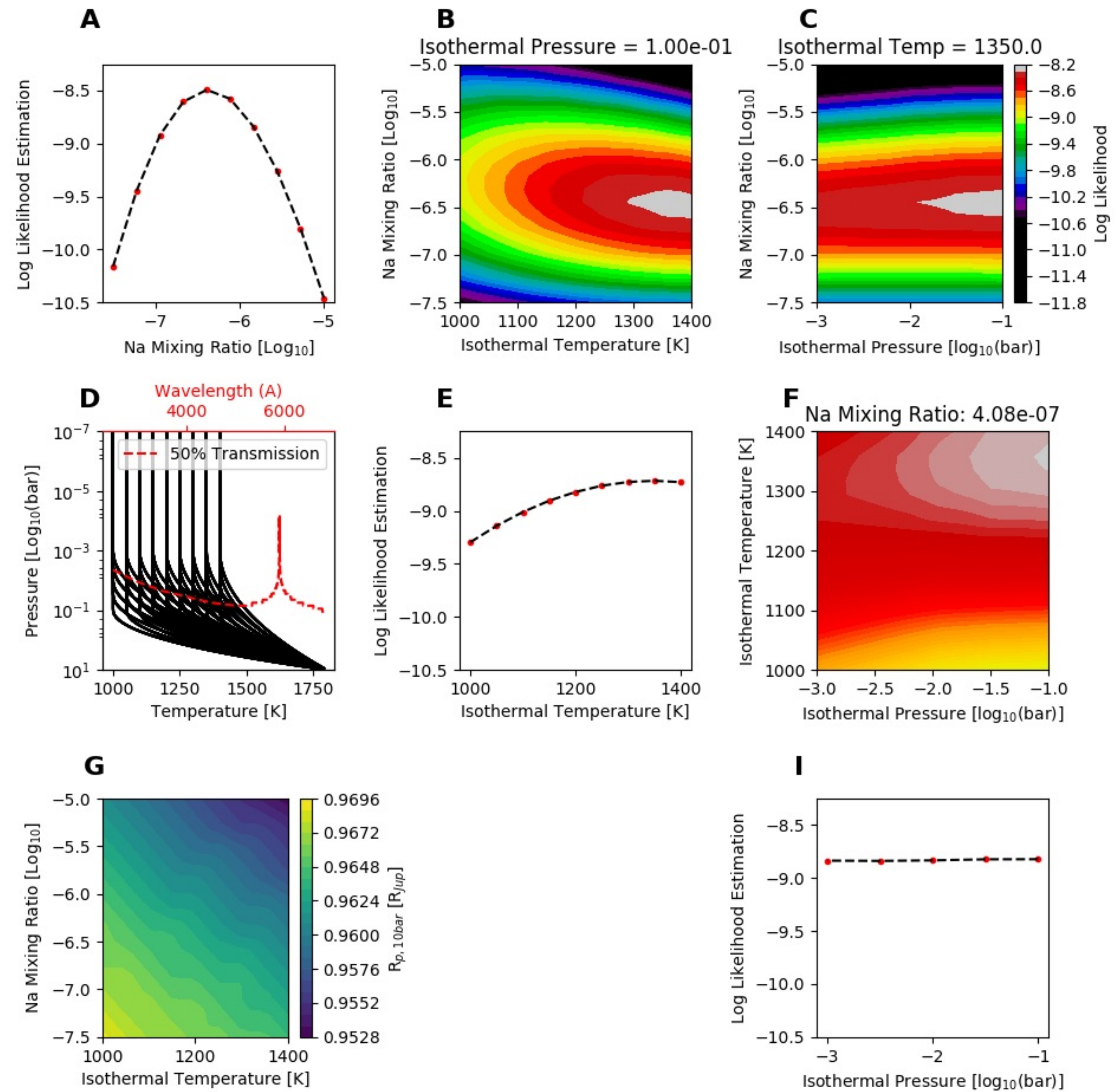}
\caption[XO-2b Grid Search]{ Likelihood estimations and correlation plots are calculated from our RT grid search. \textbf{B},\textbf{C} and \textbf{F} are correlation plots constructed from cross sections of our grid search. The optimal cross section is reported in the title and taken from the parameter set with the maximum log likelihood. Plots \textbf{A,E,I} are averaged log-likelihood estimates for each parameter at the cross section given in \textbf{B,C,F}. The red dots represent parameter values evaluated in our grid search with the dotted line being an interpolation. \textbf{D} shows the range of TP profiles computed from our grid search along with the level of $50\%$ transmission (from the best parameter set) as a reference for where the data roughly probes. We adjust the 10-bar radius for each parameter set to best match the data because raising the temperature at a given pressure level will increase the continuum. \textbf{G} shows the 10-bar radius value for the two parameters that vary the most, Na and isothermal temperature. Our likelihood is estimated as the exponential of the negative chi-squared.} 
\label{cloudfree_search}
\end{figure*}

\begin{figure*}[ht]
\centering
\includegraphics[scale=0.80]{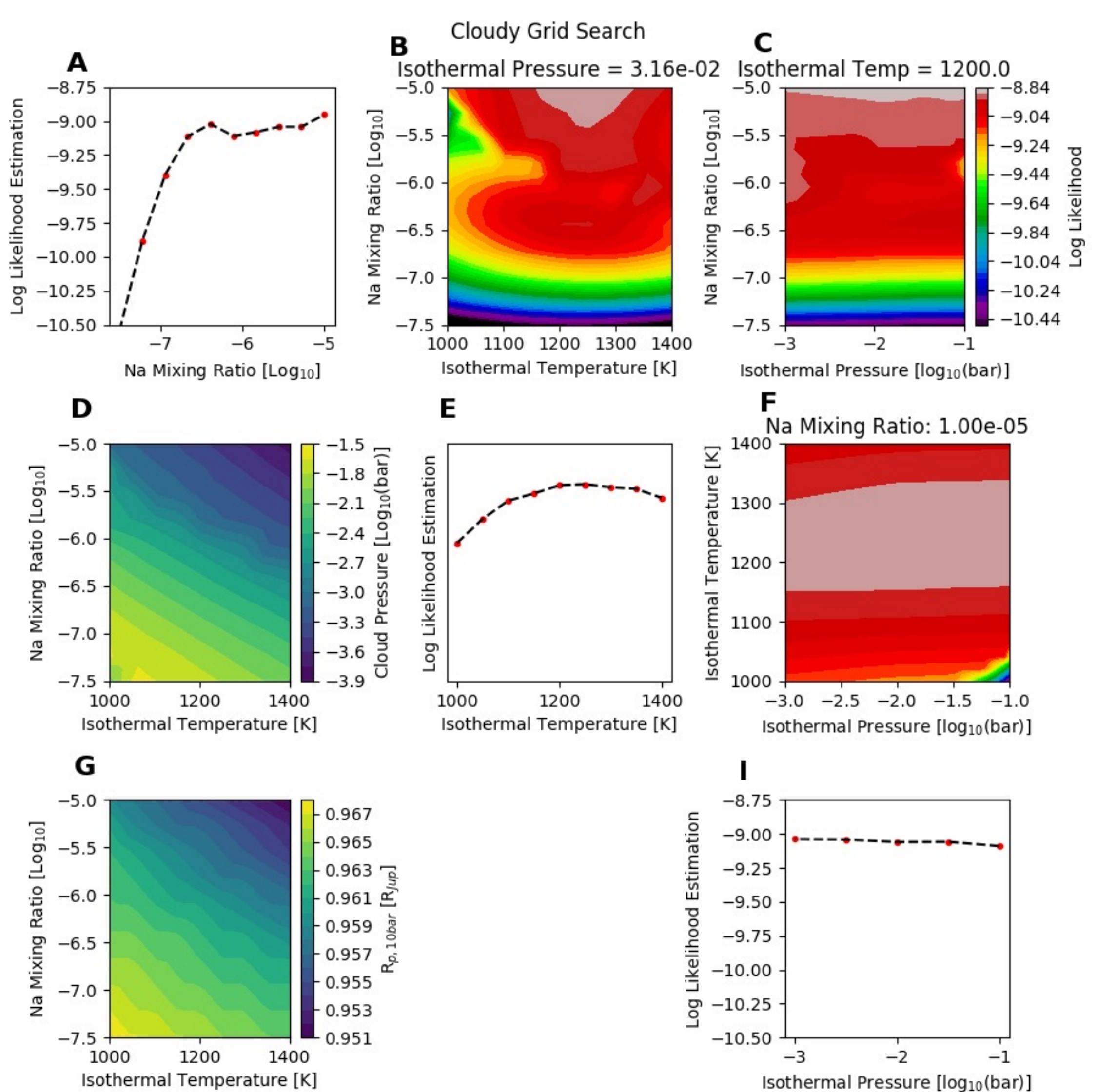}
\caption[XO-2b Grid Search]{ Likelihood estimations and correlation plots are calculated from our RT grid search. \textbf{B},\textbf{C} and \textbf{F} are correlation plots constructed from cross sections of our grid search. The optimal cross section is reported in the title and taken from the parameter set with the maximum log likelihood. Plots \textbf{A,E,I} are averaged log-likelihood estimates for each parameter at the cross section given in \textbf{B,C,F}. The red dots represent parameter values evaluated in our grid search with the dotted line being an interpolation. \textbf{D} shows the injected cloud top pressure in relation to Na abundance and atmospheric temperature. We adjust the 10-bar radius for each parameter set to best match the data because raising the temperature at a given pressure level will increase the continuum. \textbf{G} shows the 10-bar radius value for the two parameters that vary the most, Na and isothermal temperature. Our likelihood is estimated as the exponential of the negative chi-squared. We find clouds introduce a degenerate solution set with similar fits by increasing the Na abundance while simultaneously decreasing the 10-bar radius, increasing the atmospheric temperature and injecting a cloud to mask the Rayleigh scattering slope. } 
\label{cloudy_search}
\end{figure*}

The free variables in our radiative transfer model are the initial mixing ratio of Na and a 2-parameter temperature profile. The 10-bar radius is optimized between 0.95 -- 0.97 $R_{Jup}$ to fit the continuum, which is sensitive to the temperature. We vary the Na abundance on a uniform Log$_{10}$ scale with 10 points between 10$^{-5}$ to 10$^{-7.5}$ ppm. The temperature structure is modeled with an isothermal temperature in the upper atmosphere and a transition pressure, below which the atmosphere warms following a dry adiabat to 1800 K at 10-bar (Figure \ref{atmo_model}). We adopt a 10-bar temperature of 1800 K, based on terminator average conditions for HD 209458 b, a hot-Jupiter with properties similar to XO-2 b \citep{Showman2009}. The isothermal temperatures range from 1000 to 1400 K at increments of 50 K, while the transition pressure ranges between 1 and 100 mbar on a log pressure scale with 5 points (Figure \ref{transmission_comparison}).

The shortwave data, outside the Na feature, establishes the continuum of the spectrum, which depends on the effects of possible clouds, the planet’s radius, as well as the temperature profile, through the scale height. The slope of the transit depth at optical wavelengths ($<\sim$5300 $\AA$) is sensitive to presence of clouds in an exoplanet atmosphere. Optically thick clouds truncate the slope longward of a certain wavelength depending on the pressure level of the cloud. Whereas a cloud-free atmosphere at the observable level ($\sim$1--100 mbar) displays a Rayleigh slope consistent with the known scale height of the atmosphere. We run an additional grid search over the same parameters as above, with the exception that in each evaluation we inject a cloud and optimize the pressure level. We consider only optically thick cloud decks ranging from $\sim$1 to 100 mbar, which roughly spans the pressures probed by our data (see Figure \ref{atmo_model}). This cloud model is parameterized simply by a cloud top pressure such that the transmission is zero at pressures greater than the cloud top. 

\textcolor{black}{We adopt different wavelength resolutions in figure \ref{transmission_comparison} to optimize the signal for spectral variations in each wavelength region. The blue region experiences smaller amplitude variation due to scattering than the Na feature and smaller variation with respect to wavelength. Therefore, only a few data points are required to constrain the blue side of the data and we can optimie for signal with larger bins. Defining the Na line profile requires higher resolution and thus we optimize for wavelength resolution but compromise SNR. }

\section{Results}

Because of our wavelength analysis, the derived spectra of XO-2b from Gemini and Kitt Peak data indicate consistent results. Light curves from both GMOS and KOSMOS exhibit absorption at 589nm indicative of Na. The wavelength calibration resolves the combined absorption of the Na doublet with $\sim$16$\AA$ wide bins. The absorption depth of the Na I feature is determined to 3.8 $\sigma$ using both datasets, and the average spectral amplitude is 459$\pm$120 ppm. The individual amplitudes of the Na feature are 521$\pm$161 ppm and 403$\pm$186 ppm for GMOS and KOSMOS respectively. The amplitude of the Na feature extends to $\sim$11 scale heights above an upper limit to the 10-bar radius of 0.962 R$_{jup}$. 

\subsection{Atmospheric Signatures}

Our analysis of the data assuming cloudy and cloud-free atmospheres yields similar results, because the shortwave data indicates the continuum, and the 10 bar radius is therefore adjusted to fit the continuum. The transition pressure, isothermal temperature and Na mixing ratio with the highest likelihood is 100 mbar, 1350K and 4.08e-7 for the clear atmosphere, and 31.6 mbar, 1200K and 1.0e-5 for the cloudy atmosphere assuming a 10 bar radius of 0.961 R$_{Jup}$ with a cloud deck at 2.4 mbar (See Figure \ref{cloudfree_search} and \ref{cloudy_search}). The BIC value of our ``Cloud Free'' fit is 20 while the BIC of the cloudy model is 25. It's interesting to note that the ``Cloud Free'' and ``Cloudy'' models have the same chi-squared value but different number of parameters (e.g. 3 and 4, respectively) and thus different BICs. Our data lacks the signal to noise to discriminate the presence of clouds and hazes between $\sim$1-100 mbar (see Figure \ref{transmission_comparison}). 

We find that changes in the temperature at 10 bar $\pm$200 K had a negligible effect on our model fits. Therefore, our assumption to use 1800 K at 10 bar is adequate. Interpretation of our transit measurements do not depend sensitively on the pressure of our parameterized TP profile either. However, there is a small dependence on the temperature within our assumed isothermal portion of the atmosphere (see Figure \ref{cloudfree_search}). We find an isothermal temperature of 1350 K best fits our data and it is consistent with the equilibrium temperature and our estimation from the Rayleigh slope. As the temperature goes up, the density at a particular pressure level increases contributing to a larger transit depth. Despite the density of Na increasing in this manner, the photoionization level in the atmosphere is weakly dependent on the temperature.

We derive a lower limit to the Na abundance at 0.4$^{+2}_{-0.3}$ ppm ([Na/H]=-0.64$^{+0.78}_{-0.6}$), which is consistent with solar values, and note that the inclusion of the effects of ionization affect the derived Na abundance \citep{Asplund2009}. However, Our derived Na abundances fall short of the metallicity ([Na/H]) of XO-2N of 0.485 $\pm$ 0.043, which indicate an enhancement of Na compared to solar values (\citealt{Biazzo2015}; \citealt{Teske2015}). At pressures smaller than 0.1 mbar Na is ionized in the upper atmosphere and no longer contributes to the optical depth at visible wavelengths. The Na abundance is constrained by the amplitude of our feature relative to the wings. While the wings of the feature are sensitive to abundance at a given temperature, we find that the amplitude actually decreases with increasing abundance. As the abundance increases, the level of the band wings increases faster than the peak, due to the peak being set by ionization. \textcolor{black}{In this case, the Na line profile is weakly constrained by the data.} However, it is possible to achieve a larger Na abundance by introducing a cloud, lowering the 10-bar radius and increasing the temperature to counteract the lower continuum. However, this would require masking the Rayleigh scattering slope with a cloud, creating a degenerate solution set between 10-bar radius, atmospheric temperature, cloud pressure and Na abundance profile. Figure \ref{cloudy_search} shows this degeneracy since we are only able to place a lower limit on the derived Na abundance at 0.4$^{+2}_{-0.3}$ ppm. Our data are consistent with a clear atmosphere between $\sim$1--100 mbar which allow us to constrain the Na abundance. However, we can not rule out the presence of clouds at $\sim$10 mbar which would make our results consistent with the stellar Na metallicity but introduce additional degeneracies (see Figure \ref{cloudy_search})

Our simplistic photoionization model is consistent with a more complicated photochemical model for XO-2b that includes K as an electron source and excited state chemistry for Na and K \citep{Lavvas2014}. The ionization potential for Na is low enough for thermal ionization to play a large role in creating the electron densities low in the atmosphere \citep{Koskinen2014}. However, XO-2 b is not hot enough for this to be a dominant ion producing mechanism (see Figure \ref{atmo_model}). The EUV flux reaches as far as 1 bar in our atmospheric model however the ion density does not dominate until 0.1 mbar consistent with \citep{Lavvas2014}. We find that the dominate mechanisms necessary to reproduce consistent results to within our observable range are photoionization, radiative recombination and 3-body recombination.

Our derived absorption spectrum differs from that of the first detection of Na in the atmosphere of XO-2b, mainly because we have information on the band wings, and the continuum. This previous detection of Na was influenced by seeing-induced slit loses and only a comparison spectrum (where the average transit depth is divided out of the data) was created \citep{Sing2012}. \citealt{Sing2012} were unable to resolve the band wings in 5 nm wide bins potentially due to the presence of obscuring hazes that are below the pressure level of where Na resides. 

\subsection{Rayleigh Scattering and Clouds}

We test our radiative transfer model for consistency using a simplified atmospheric calculation based on the slope of the shortwave data. Assuming an estimated mean molecular weight and gravity, the detection of a Rayleigh scattering slope yields the atmospheric scale height and thus empirically determines the planet's temperature at the day-night terminator. The data short ward of 5300 $\AA$ defines the Rayleigh slope (where $\tau$ $>>$ 1). The slope of the transmission spectrum is linearly proportional to the scattering index and scale height where the atmosphere opacity exhibits a power law dependence with respect to wavelength. The temperature is determined by

\begin{equation} \label{rayleigh}
\alpha T = \frac{\mu g }{k} \frac{d R_{p}}{d \ln{\lambda} }
\end{equation}

\noindent \citep{Lecavelier2008}. Here $\mu$ is the mean molecular mass, which we estimate to be 2.3 m$_{amu}$ given an atmosphere dominated by H$_{2}$ and He; $g$ is the surface gravity with a value of 1523 $cm/s^{2}$ (derived from our 10-bar radius); $k$ is the Boltzmann constant; $T$ is temperature and, $\alpha$ is the index that defines the wavelength dependence of the scattering cross-section, $\sigma/\sigma_{0}$ = $(\lambda/\lambda_{0})^{\alpha}$. For Rayleigh scattering $\alpha$=-4 \citep{Hansen1974}. To compute the radius of the planet we multiply our $R_{p}/R_{*}$ measurements by a stellar radius value of 0.971 $R_{sun}$ \citep{Torres2008}. The slope of our data was determined using a weighted least squares fit in semi-logarithmic space where only the wavelengths were in log values. We find that our slope is consistent with an atmospheric temperature of 1450K, which is within the range of potential equilibrium temperatures for XO-2b, 1046 -- 1361 K, and our best fit isothermal temperature, 1350 K. It is unlikely for optically thick clouds to be at altitudes higher than $\sim$1 mbar because they would flatten the Rayleigh slope causing this temperature estimate to decrease, consistent with a flat spectrum. 

\textcolor{black}{ Measurements of the Rayleigh slope with more than one instrument is best done simultaneously or at similar stellar activity phases since star spots can influence the transit depth (\citealt{Mccullough2014}; \citealt{Zellem2017}). \cite{Zellem2015} indicates that XO-2N is variable, potentially due to cool star spots, with a peak-to-peak amplitude of 0.0049 $\pm$ 0.0007 R-mag and a period of 29.89 $\pm$ 0.16 days for the 2013–2014 observing season and a peak-to-peak amplitude of 0.0035 $\pm$ 0.0007 R-mag and 27.34 $\pm$ 0.21 day period for the 2014–2015 observing season. At worst, the star varies by 0.45$\%$ which would correspond to the transit depth changing by $\sim$48 ppm (using equation 7 in \citealt{Zellem2017}). This effect is 8.2 times less than the amplitude of the Na feature. Therefore, we can assume that it is not induced by stellar activity. The transit depth uncertainties are between 200-500 ppm so a perturbation of $\sim$50 ppm is not distinguishable from the noise. Additionally, the GMOS observations were conducted at a stellar phase of 0.07 and the phase for the Kitt Peak observations was 0.96. A phase difference of 0.11 between the two observations corresponds to a three day difference in phase space on the rotation of XO-2N. In conclusion, despite XO-2N being an active star we took measures to observe the system at similar stellar phases and in the worst case scenario, if the star was active, the influence would be indistinguishable from the noise. 
}

\section{Conclusion and Future Work}

Here we present new observations and analyses of the Na abundance of the exoplanet XO-2 b through transit spectroscopy recorded at the Gemini/GMOS and Mayall/KOSMOS telescopes. We find that the astrophysical signals are subject to time-varying translations along the detector that change according to wavelength. A cross-correlation in both time and wavelength are used to correct for misalignments in the pixel-wavelength solution for each image. Improper alignment prior to dividing the astrophysical signals can result in spurious spectral features or inadequate removal of shared systematics. A quick way to diagnose a non-linear misalignment in two spectra is to cross-correlate each side of the detector and check that the peak of the phase only correlation function yields the same value (See figure \ref{obs_cc}). 

Exoplanets orbiting close to their host star are subject to ionizing radiation that will change the abundance of neutral constituents as a function of altitude. We couple a photoionization and radiative transfer model to interpret our transit measurements. The amplitude of Na absorption is 459$\pm$120 ppm corresponding to limit on the Na mixing ratio of 0.4$^{+2}_{-0.3}$ ppm. The data are consistent with a clear atmosphere between $\sim$1--100 mbar however we can not rule out optically thick clouds at pressures greater than $\sim$100 mbar. However, more precise measurements of the Rayleigh scattering slope could help detect the presence of hazes or constrain the effects due to stellar activity. 

Further transit observations of XO-2b will be able to constrain the K abundance in the atmosphere and place constraints on the vertical temperature profile of the planet and thus condensation regime and cloud properties. We urge caution to observers of this system as the host star and binary companion exhibit some amount of spectral variability which can perturb the transit measurements. Past photometric and spectroscopic measurements have revealed XO-2N to have a rotation period on the order of one month (\citealt{Damasso2015}; \citealt{Zellem2015}). Observing the object at similar stellar phases can minimize the difference in stellar surface features, which can perturb the signal of the planet through star spots. 

\section{Acknowledgements}
Part of the research was carried out at the Jet Propulsion Laboratory, California Institute of Technology, under contract with the National Aeronautics and Space Administration. We would like to thank Dr. Mark Swain, Dr. David Koerner and Dr. David Trilling for their support in reviewing an early portion of this research. Along with all of the telescope staff at Kitt Peak and Gemini for facilitating these observations. 

Visiting Astronomer, Kitt Peak Mayall 4m, National Optical Astronomy Observatories, which is operated by the Association of Universities for Research in Astronomy, Inc.  (AURA) under cooperative agreement with the National Science Foundation.

Based on observations obtained at the Gemini Observatory, which is operated by the Association of Universities for Research in Astronomy, Inc., under a cooperative agreement with the NSF on behalf of the Gemini partnership: the National Science Foundation (United States), the National Research Council (Canada), CONICYT (Chile), Ministerio de Ciencia, Tecnolog\'{i}a e Innovaci\'{o}n Productiva (Argentina), and Minist\'{e}rio da Ci\^{e}ncia, Tecnologia e Inova\c{c}\~{a}o (Brazil).

\singlespace
\bibliography{ref}

\begin{figure*}
    \centering
    \hspace*{-.5in}
       \includegraphics[scale=0.85]{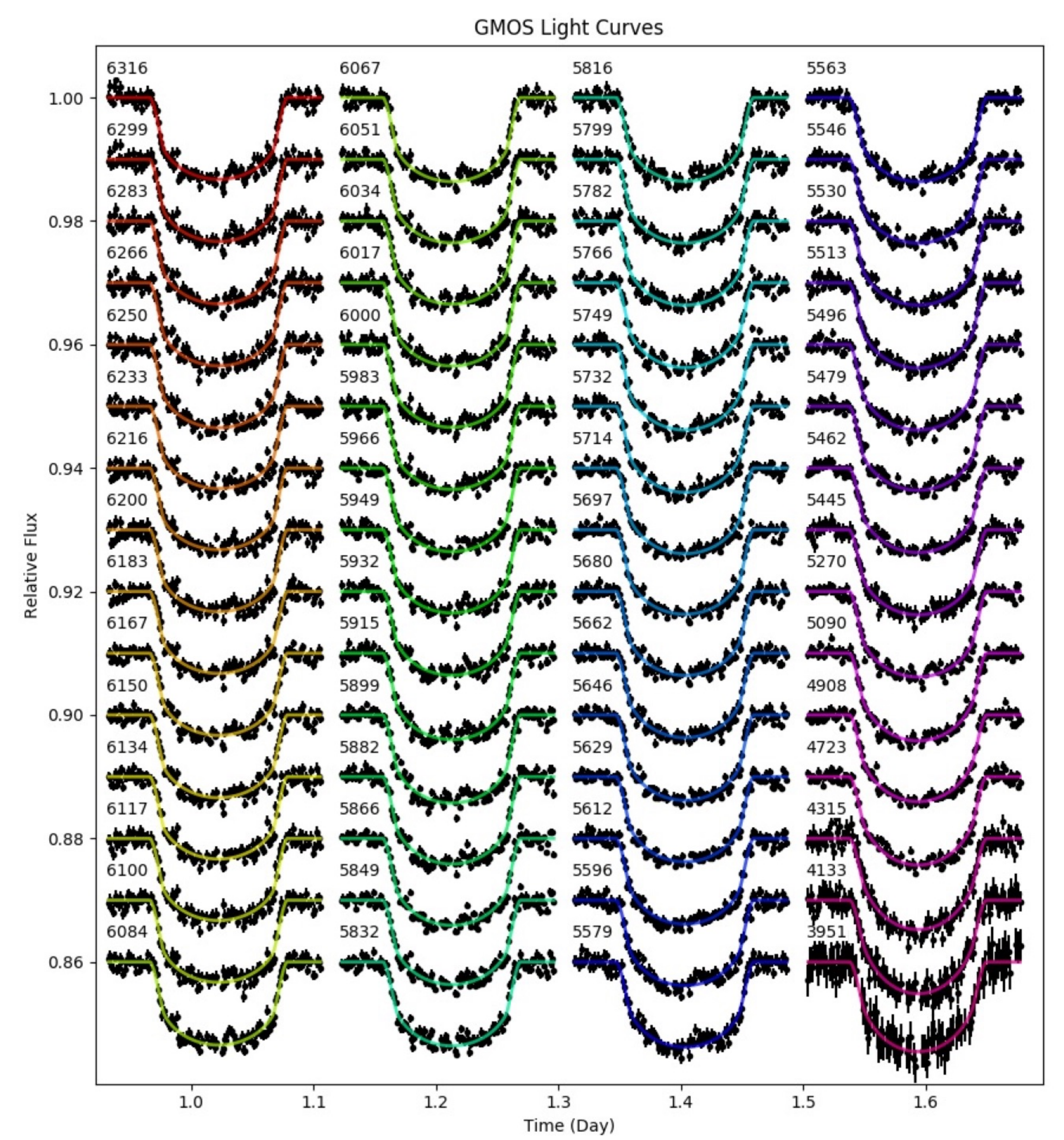}
\caption[GMOS Light Curves]{ \textcolor{black}{Data from the GMOS instrument are shown with their respective light curve model. The model was derived from the nested sampling algorithm leaving the transit depth and limb darkening parameters free along with two parameters that modeled residual curvature. Each color represents a different wavelength and the wavelengths are shown in angstroms as text above the light curve.} }
    \label{fig:a}
    \label{gmos_lc}
\end{figure*}

\begin{figure*}
    \centering
    \hspace*{-.5in}
       \includegraphics[scale=0.85]{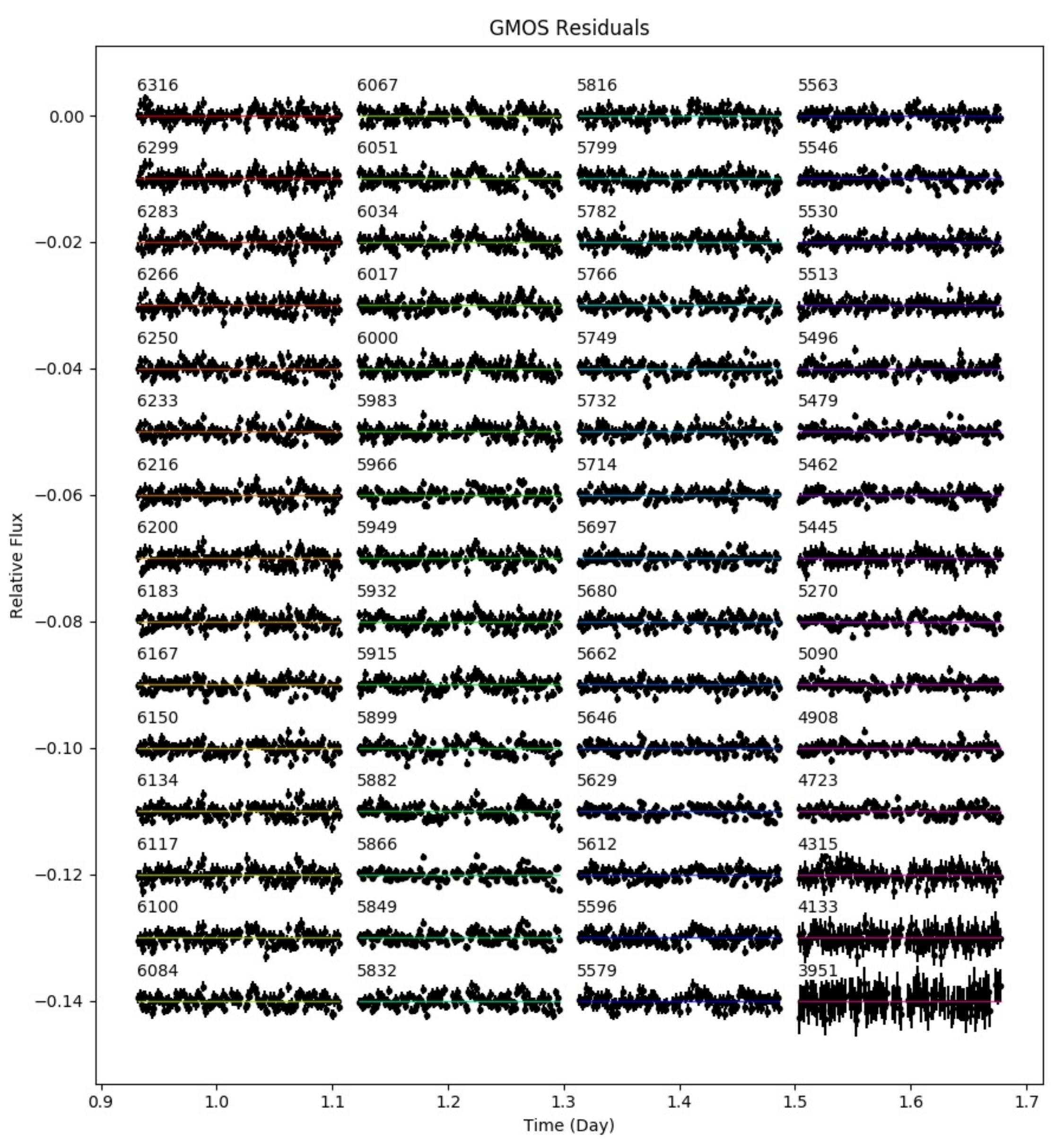}
\caption[GMOS Light Curves]{ \textcolor{black}{Residual data from the GMOS light curve models shown in Figure 13. Each color represents a different wavelength and the wavelengths are shown in angstroms as text above the light curve. }}
    \label{fig:a}
    \label{gmos_lc}
\end{figure*}

\begin{figure*}
    \centering
    \hspace*{-.5in}
       \includegraphics[scale=0.85]{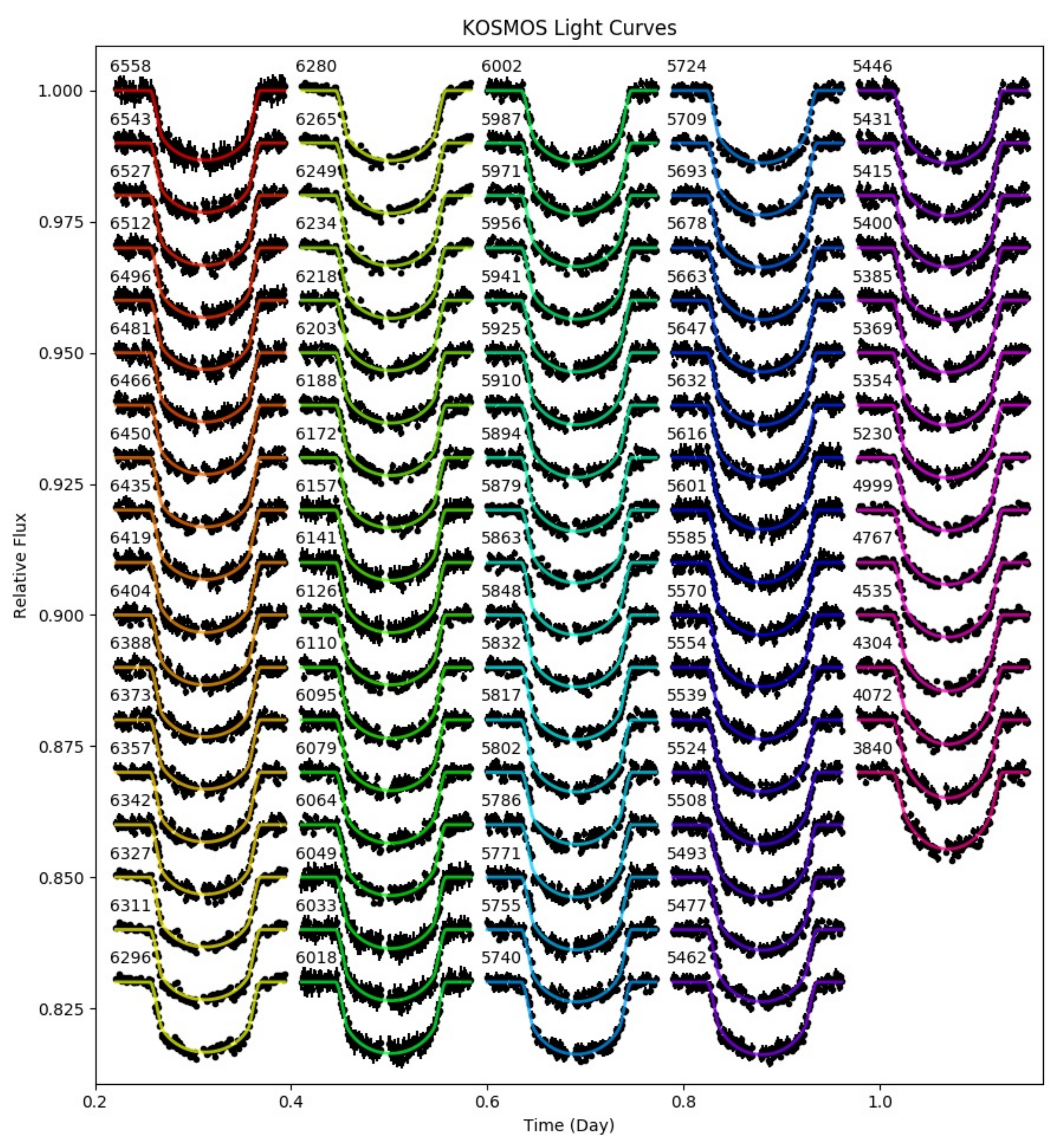}
\caption[GMOS Light Curves]{ \textcolor{black}{Data from the KOSMOS instrument are shown with their respective light curve model. The model was derived from the nested sampling algorithm leaving the transit depth and limb darkening parameters free along with two parameters that modeled residual curvature. Each color represents a different wavelength and the wavelengths are shown in angstroms as text above the light curve.} Table 3 has the final parameters for the model fit.}
    \label{fig:a}
    \label{kosmos_lc}
\end{figure*}
\begin{figure*}
    \centering
    \hspace*{-.5in}
       \includegraphics[scale=0.85]{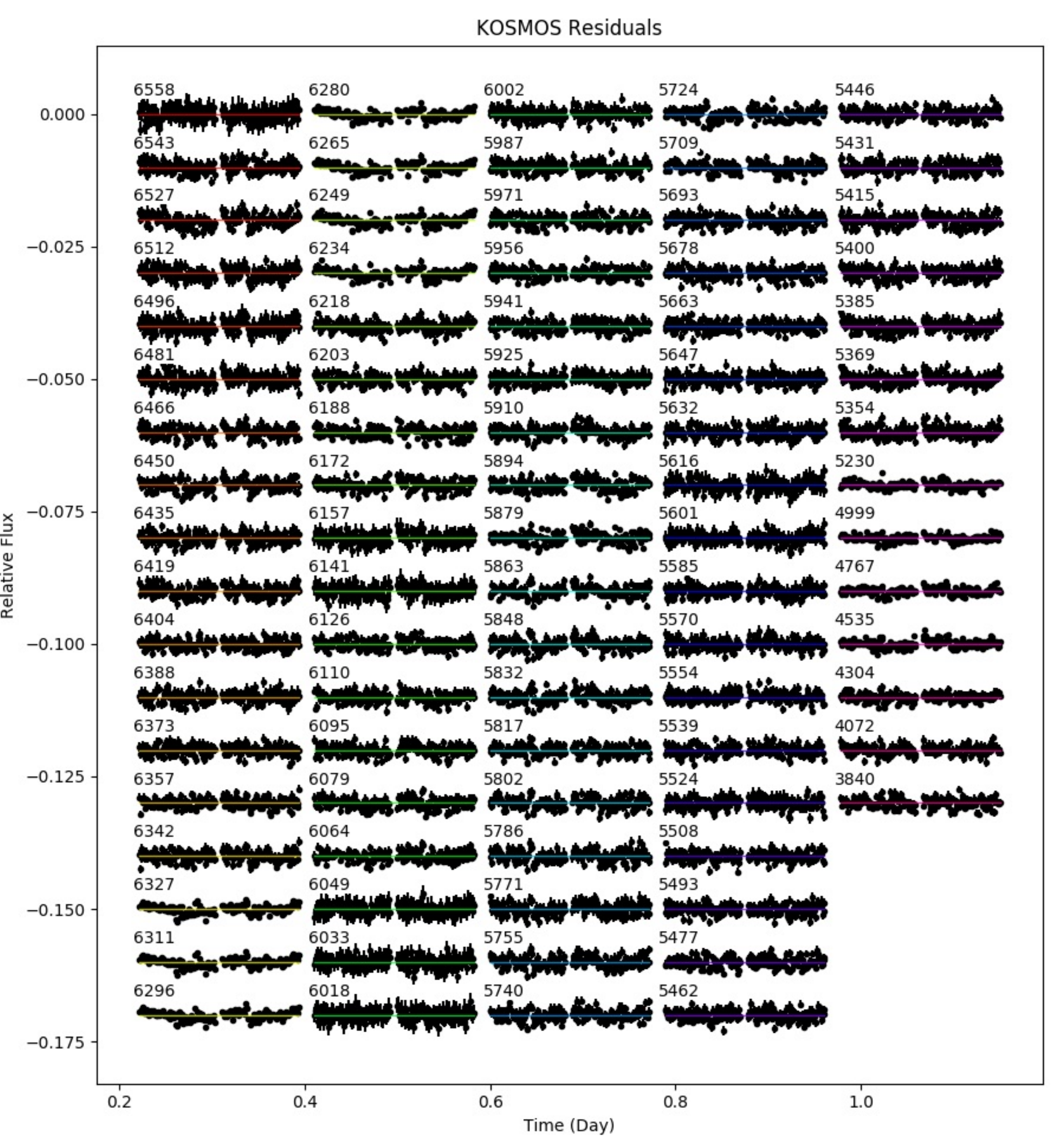}
\caption[GMOS Light Curves]{ \textcolor{black}{Residual data from the GMOS light curve models shown in Figure 13. Each color represents a different wavelength and the wavelengths are shown in angstroms as text above the light curve.} }
    \label{fig:a}
    \label{kosmos_lc}
\end{figure*}


\begin{table*}
  \caption{GMOS Final Model Parameters}
  \label{tab:finalpars}
  \begin{center}
    \leavevmode
    \begin{tabular}{c c c c c c } \hline \hline
Wavelength (A) & R$_{p}$/R$_{s}$ & u$_{1}$ & a$_{0}$ & a$_{1}$ & $\sigma_{res}$ (ppm)  \\
\hline
6308-6324 &0.103328$\pm$0.001095&0.64$\pm$0.05& 0.959$\pm$0.002&0.0104$\pm$0.0019& 1015 \\
6291-6308 &0.103694$\pm$0.001112&0.64$\pm$0.05& 0.964$\pm$0.002&0.0087$\pm$0.0019& 974 \\
6274-6291 &0.103803$\pm$0.001132&0.65$\pm$0.05& 0.968$\pm$0.002&0.0053$\pm$0.0018& 1018 \\
6258-6274 &0.104006$\pm$0.001143&0.65$\pm$0.05& 0.966$\pm$0.002&0.0055$\pm$0.0019& 1025 \\
6241-6258 &0.104155$\pm$0.001016&0.66$\pm$0.04& 0.966$\pm$0.002&0.0066$\pm$0.0018& 1015 \\
6225-6241 &0.103529$\pm$0.000920&0.66$\pm$0.04& 0.969$\pm$0.002&0.0048$\pm$0.0016& 964 \\
6208-6225 &0.103293$\pm$0.000977&0.65$\pm$0.04& 0.968$\pm$0.002&0.0075$\pm$0.0017& 995 \\
6192-6208 &0.103334$\pm$0.001108&0.65$\pm$0.05& 0.965$\pm$0.002&0.0117$\pm$0.0019& 1157 \\
6175-6192 &0.103406$\pm$0.001018&0.65$\pm$0.05& 0.964$\pm$0.002&0.0093$\pm$0.0018& 1039 \\
6158-6175 &0.103525$\pm$0.000923&0.65$\pm$0.04& 0.965$\pm$0.002&0.0068$\pm$0.0015& 891 \\
6142-6158 &0.103929$\pm$0.000942&0.65$\pm$0.04& 0.966$\pm$0.002&0.0067$\pm$0.0016& 928 \\
6125-6142 &0.103405$\pm$0.001023&0.66$\pm$0.04& 0.966$\pm$0.002&0.0070$\pm$0.0018& 961 \\
6109-6125 &0.103571$\pm$0.001078&0.65$\pm$0.05& 0.966$\pm$0.002&0.0088$\pm$0.0019& 1991 \\
6092-6109 &0.103468$\pm$0.001081&0.65$\pm$0.05& 0.971$\pm$0.002&0.0063$\pm$0.0018& 1990 \\
6075-6092 &0.103865$\pm$0.001057&0.66$\pm$0.05& 0.970$\pm$0.002&0.0099$\pm$0.0018& 951 \\
6059-6075 &0.104390$\pm$0.001097&0.66$\pm$0.05& 0.973$\pm$0.002&0.0101$\pm$0.0020& 965 \\
6042-6059 &0.104141$\pm$0.000993&0.66$\pm$0.04& 0.977$\pm$0.002&0.0066$\pm$0.0018& 998 \\
6025-6042 &0.103702$\pm$0.001039&0.67$\pm$0.05& 0.979$\pm$0.002&0.0035$\pm$0.0018& 978 \\
6008-6025 &0.103555$\pm$0.001010&0.67$\pm$0.05& 0.979$\pm$0.002&0.0028$\pm$0.0018& 924 \\
5992-6008 &0.103707$\pm$0.001089&0.66$\pm$0.05& 0.982$\pm$0.002&0.0011$\pm$0.0018& 889 \\
5975-5992 &0.103929$\pm$0.001005&0.66$\pm$0.04& 0.984$\pm$0.002&-0.0003$\pm$0.0018& 885 \\
5958-5975 &0.103849$\pm$0.000832&0.67$\pm$0.03& 0.984$\pm$0.001&-0.0015$\pm$0.0015& 863 \\
5941-5958 &0.103580$\pm$0.000949&0.68$\pm$0.04& 0.983$\pm$0.002&-0.0013$\pm$0.0017& 1004 \\
5924-5941 &0.103919$\pm$0.001008&0.68$\pm$0.04& 0.981$\pm$0.002&-0.0016$\pm$0.0017& 1038 \\
5907-5924 &0.105591$\pm$0.000914&0.68$\pm$0.04& 0.973$\pm$0.002&-0.0009$\pm$0.0016& 972 \\
5890-5907 &0.106607$\pm$0.001066&0.68$\pm$0.06& 0.973$\pm$0.001&-0.0002$\pm$0.0015& 994 \\
5874-5890 &0.105770$\pm$0.000964&0.70$\pm$0.04& 0.976$\pm$0.002&0.0019$\pm$0.0016& 981 \\
5857-5874 &0.105538$\pm$0.000739&0.70$\pm$0.03& 0.976$\pm$0.001&0.0011$\pm$0.0014& 840 \\
5841-5857 &0.104186$\pm$0.000869&0.69$\pm$0.04& 0.976$\pm$0.001&0.0015$\pm$0.0015& 874 \\
5824-5841 &0.103756$\pm$0.000944&0.69$\pm$0.04& 0.970$\pm$0.002&0.0065$\pm$0.0016& 958 \\
5807-5824 &0.103791$\pm$0.001097&0.69$\pm$0.05& 0.968$\pm$0.002&0.0066$\pm$0.0018& 943 \\
5791-5807 &0.103810$\pm$0.001023&0.69$\pm$0.04& 0.968$\pm$0.002&0.0058$\pm$0.0018& 839 \\
5774-5791 &0.103700$\pm$0.001094&0.70$\pm$0.05& 0.969$\pm$0.002&0.0039$\pm$0.0018& 941 \\
5758-5774 &0.103784$\pm$0.000885&0.72$\pm$0.04& 0.968$\pm$0.002&0.0049$\pm$0.0016& 930 \\
5741-5758 &0.104045$\pm$0.000896&0.73$\pm$0.04& 0.971$\pm$0.002&0.0006$\pm$0.0016& 914 \\
5723-5741 &0.104654$\pm$0.000917&0.71$\pm$0.04& 0.969$\pm$0.002&-0.0013$\pm$0.0015& 846 \\
5706-5723 &0.104536$\pm$0.000843&0.71$\pm$0.04& 0.961$\pm$0.001&0.0049$\pm$0.0014& 780 \\
5688-5706 &0.104023$\pm$0.000827&0.70$\pm$0.03& 0.964$\pm$0.001&0.0041$\pm$0.0014& 849 \\
5671-5688 &0.103509$\pm$0.000906&0.71$\pm$0.04& 0.968$\pm$0.002&-0.0007$\pm$0.0015& 835 \\
5654-5671 &0.103410$\pm$0.000892&0.72$\pm$0.04& 0.966$\pm$0.002&0.0009$\pm$0.0015& 795 \\
5637-5654 &0.104333$\pm$0.000815&0.71$\pm$0.03& 0.968$\pm$0.001&-0.0004$\pm$0.0014& 765 \\
5621-5637 &0.104194$\pm$0.000744&0.71$\pm$0.03& 0.971$\pm$0.001&-0.0021$\pm$0.0013& 744 \\
5604-5621 &0.104302$\pm$0.000951&0.72$\pm$0.04& 0.966$\pm$0.002&-0.0002$\pm$0.0016& 799 \\
5588-5604 &0.104251$\pm$0.001025&0.71$\pm$0.04& 0.964$\pm$0.002&0.0008$\pm$0.0017& 900 \\
5571-5588 &0.103961$\pm$0.001103&0.72$\pm$0.05& 0.967$\pm$0.002&0.0020$\pm$0.0019& 953 \\
5554-5571 &0.103277$\pm$0.000917&0.72$\pm$0.04& 0.968$\pm$0.002&0.0029$\pm$0.0016& 813 \\
5538-5554 &0.103333$\pm$0.000869&0.71$\pm$0.03& 0.968$\pm$0.002&0.0027$\pm$0.0015& 834 \\
5521-5538 &0.103331$\pm$0.000930&0.72$\pm$0.04& 0.967$\pm$0.002&0.0014$\pm$0.0016& 772 \\
5505-5521 &0.103921$\pm$0.001001&0.73$\pm$0.04& 0.970$\pm$0.002&-0.0011$\pm$0.0018& 784 \\
5488-5505 &0.104182$\pm$0.000904&0.72$\pm$0.04& 0.968$\pm$0.002&0.0019$\pm$0.0016& 923 \\
5471-5488 &0.103268$\pm$0.000825&0.72$\pm$0.04& 0.966$\pm$0.001&0.0031$\pm$0.0013& 859 \\
5454-5471 &0.103260$\pm$0.000899&0.74$\pm$0.04& 0.969$\pm$0.002&0.0007$\pm$0.0015& 777 \\
5437-5454 &0.103611$\pm$0.001198&0.74$\pm$0.05& 0.969$\pm$0.002&-0.0004$\pm$0.0020& 898 \\
5178-5363 &0.104007$\pm$0.000809&0.73$\pm$0.04& 0.960$\pm$0.001&-0.0013$\pm$0.0014& 797 \\
5002-5178 &0.103971$\pm$0.000791&0.79$\pm$0.03& 0.951$\pm$0.001&0.0000$\pm$0.0013& 730 \\
4814-5002 &0.103252$\pm$0.000788&0.80$\pm$0.03& 0.942$\pm$0.001&0.0051$\pm$0.0014& 883 \\
4633-4814 &0.103371$\pm$0.000820&0.84$\pm$0.03& 0.933$\pm$0.001&0.0045$\pm$0.0014& 1048 \\
4224-4405 &0.104040$\pm$0.001518&0.88$\pm$0.06& 0.908$\pm$0.002&-0.0006$\pm$0.0024& 1239 \\
4042-4224 &0.104290$\pm$0.001888&0.92$\pm$0.07& 0.906$\pm$0.003&0.0008$\pm$0.0030& 1393 \\
3860-4042 &0.103500$\pm$0.003212&0.87$\pm$0.12& 0.897$\pm$0.005&-0.0006$\pm$0.0050& 3317 \\
    \end{tabular}
  \end{center}
\end{table*}

\begin{table*}
  \caption{KOSMOS Final Model Parameters (I)}
  \label{tab:finalpars2}
  \begin{center}
    \leavevmode
    \begin{tabular}{c c c c c c } \hline \hline
Wavelength (A) & R$_{p}$/R$_{s}$ & u$_{1}$ & a$_{0}$ & a$_{1}$ & $\sigma_{res}$ (ppm)  \\
\hline
6550-6566 &0.103695$\pm$0.001434&0.64$\pm$0.06& 1.016$\pm$0.001&-0.0097$\pm$0.0026& 885 \\
6535-6550 &0.103917$\pm$0.000990&0.62$\pm$0.04& 0.976$\pm$0.001&0.0050$\pm$0.0017& 898 \\
6520-6535 &0.104135$\pm$0.001033&0.63$\pm$0.04& 0.976$\pm$0.001&0.0080$\pm$0.0018& 972 \\
6504-6520 &0.103679$\pm$0.001101&0.64$\pm$0.05& 0.975$\pm$0.001&0.0046$\pm$0.0019& 937 \\
6489-6504 &0.103779$\pm$0.001227&0.62$\pm$0.05& 0.974$\pm$0.001&0.0025$\pm$0.0021& 966 \\
6473-6489 &0.103745$\pm$0.001159&0.62$\pm$0.05& 0.969$\pm$0.001&0.0069$\pm$0.0020& 915 \\
6458-6473 &0.103743$\pm$0.001007&0.62$\pm$0.04& 0.969$\pm$0.001&0.0065$\pm$0.0017& 895 \\
6442-6458 &0.103786$\pm$0.000944&0.62$\pm$0.05& 0.969$\pm$0.000&0.0074$\pm$0.0016& 854 \\
6427-6442 &0.103613$\pm$0.001037&0.63$\pm$0.04& 0.972$\pm$0.001&0.0039$\pm$0.0018& 878 \\
6412-6427 &0.104080$\pm$0.001011&0.63$\pm$0.04& 0.972$\pm$0.001&0.0059$\pm$0.0017& 914 \\
6396-6412 &0.104029$\pm$0.000986&0.63$\pm$0.04& 0.974$\pm$0.001&0.0013$\pm$0.0017& 881 \\
6381-6396 &0.103681$\pm$0.001038&0.63$\pm$0.04& 0.972$\pm$0.001&0.0017$\pm$0.0018& 839 \\
6365-6381 &0.103638$\pm$0.000916&0.63$\pm$0.04& 0.970$\pm$0.000&0.0025$\pm$0.0016& 835 \\
6350-6365 &0.103933$\pm$0.000774&0.63$\pm$0.03& 0.969$\pm$0.000&0.0052$\pm$0.0013& 823 \\
6334-6350 &0.103695$\pm$0.000894&0.63$\pm$0.04& 0.968$\pm$0.000&0.0053$\pm$0.0015& 882 \\
6319-6334 &0.103831$\pm$0.000559&0.63$\pm$0.02& 0.973$\pm$0.000&0.0010$\pm$0.0009& 705 \\
6303-6319 &0.104105$\pm$0.000571&0.63$\pm$0.02& 0.972$\pm$0.000&0.0010$\pm$0.0009& 703 \\
6288-6303 &0.103957$\pm$0.000598&0.64$\pm$0.02& 0.972$\pm$0.000&0.0015$\pm$0.0010& 725 \\
6273-6288 &0.103856$\pm$0.000638&0.64$\pm$0.03& 0.971$\pm$0.000&0.0030$\pm$0.0011& 721 \\
6257-6273 &0.104119$\pm$0.000699&0.64$\pm$0.03& 0.970$\pm$0.000&0.0043$\pm$0.0012& 752 \\
6242-6257 &0.104231$\pm$0.000700&0.64$\pm$0.03& 0.971$\pm$0.000&0.0002$\pm$0.0012& 753 \\
6226-6242 &0.104280$\pm$0.000764&0.64$\pm$0.03& 0.971$\pm$0.000&-0.0001$\pm$0.0013& 755 \\
6211-6226 &0.104154$\pm$0.000928&0.64$\pm$0.04& 0.969$\pm$0.000&0.0071$\pm$0.0016& 834 \\
6195-6211 &0.104015$\pm$0.000935&0.65$\pm$0.04& 0.967$\pm$0.001&0.0042$\pm$0.0016& 831 \\
6180-6195 &0.104097$\pm$0.000785&0.65$\pm$0.03& 0.966$\pm$0.000&0.0066$\pm$0.0013& 794 \\
6164-6180 &0.103920$\pm$0.000895&0.65$\pm$0.04& 0.965$\pm$0.000&0.0043$\pm$0.0015& 818 \\
6149-6164 &0.104118$\pm$0.001265&0.64$\pm$0.06& 0.966$\pm$0.001&0.0051$\pm$0.0021& 862 \\
6134-6149 &0.103956$\pm$0.001338&0.64$\pm$0.06& 0.968$\pm$0.001&0.0009$\pm$0.0023& 845 \\
6118-6134 &0.103749$\pm$0.000880&0.64$\pm$0.04& 0.968$\pm$0.000&0.0022$\pm$0.0014& 849 \\
6103-6118 &0.104324$\pm$0.000881&0.65$\pm$0.04& 0.971$\pm$0.000&0.0010$\pm$0.0015& 803 \\
6087-6103 &0.104213$\pm$0.000902&0.65$\pm$0.04& 0.970$\pm$0.001&0.0053$\pm$0.0016& 833 \\
6072-6087 &0.104353$\pm$0.000823&0.65$\pm$0.03& 0.970$\pm$0.000&0.0048$\pm$0.0015& 820 \\
6056-6072 &0.104063$\pm$0.000771&0.66$\pm$0.03& 0.970$\pm$0.000&0.0020$\pm$0.0014& 814 \\
6041-6056 &0.104454$\pm$0.001349&0.66$\pm$0.05& 0.970$\pm$0.001&0.0020$\pm$0.0024& 862 \\
6025-6041 &0.104208$\pm$0.001254&0.66$\pm$0.05& 0.969$\pm$0.001&0.0016$\pm$0.0022& 902 \\
6010-6025 &0.104325$\pm$0.001350&0.67$\pm$0.06& 0.967$\pm$0.001&0.0045$\pm$0.0024& 839 \\
5995-6010 &0.104235$\pm$0.000991&0.67$\pm$0.04& 0.968$\pm$0.001&0.0029$\pm$0.0017& 832 \\
5979-5995 &0.103939$\pm$0.000952&0.67$\pm$0.04& 0.968$\pm$0.001&0.0033$\pm$0.0016& 864 \\
5964-5979 &0.104005$\pm$0.000872&0.67$\pm$0.04& 0.968$\pm$0.000&0.0039$\pm$0.0015& 867 \\
5948-5964 &0.104152$\pm$0.000891&0.67$\pm$0.04& 0.968$\pm$0.000&0.0024$\pm$0.0014& 840 \\
5933-5948 &0.104312$\pm$0.000905&0.67$\pm$0.04& 0.967$\pm$0.000&0.0027$\pm$0.0015& 888 \\
5917-5933 &0.104556$\pm$0.000994&0.68$\pm$0.04& 0.962$\pm$0.001&0.0005$\pm$0.0017& 819 \\
5902-5917 &0.104775$\pm$0.000894&0.68$\pm$0.04& 0.960$\pm$0.000&-0.0013$\pm$0.0015& 816 \\
5886-5902 &0.105801$\pm$0.000866&0.68$\pm$0.04& 0.958$\pm$0.000&0.0007$\pm$0.0015& 908 \\
5871-5886 &0.104947$\pm$0.000711&0.69$\pm$0.03& 0.963$\pm$0.000&0.0011$\pm$0.0012& 854 \\
5856-5871 &0.104646$\pm$0.000916&0.69$\pm$0.04& 0.963$\pm$0.000&0.0033$\pm$0.0015& 843 \\
5840-5856 &0.104328$\pm$0.000931&0.69$\pm$0.04& 0.964$\pm$0.001&0.0020$\pm$0.0016& 811 \\
5825-5840 &0.104519$\pm$0.000876&0.69$\pm$0.04& 0.964$\pm$0.000&0.0008$\pm$0.0014& 840 \\
5809-5825 &0.104416$\pm$0.000909&0.69$\pm$0.04& 0.964$\pm$0.000&-0.0038$\pm$0.0015& 840 \\
5794-5809 &0.104370$\pm$0.000977&0.69$\pm$0.04& 0.961$\pm$0.000&0.0042$\pm$0.0017& 918 \\
5778-5794 &0.104451$\pm$0.001093&0.70$\pm$0.05& 0.960$\pm$0.001&0.0043$\pm$0.0019& 905 \\
5763-5778 &0.104733$\pm$0.001162&0.70$\pm$0.05& 0.961$\pm$0.001&0.0051$\pm$0.0019& 951 \\
5747-5763 &0.104315$\pm$0.000952&0.70$\pm$0.04& 0.960$\pm$0.000&0.0030$\pm$0.0016& 893 \\
5732-5747 &0.104228$\pm$0.000855&0.70$\pm$0.04& 0.959$\pm$0.000&0.0015$\pm$0.0015& 805 \\
5717-5732 &0.104357$\pm$0.000839&0.70$\pm$0.04& 0.957$\pm$0.000&0.0041$\pm$0.0013& 853 \\
5701-5717 &0.104124$\pm$0.000848&0.70$\pm$0.04& 0.956$\pm$0.000&0.0046$\pm$0.0014& 765 \\
5686-5701 &0.104137$\pm$0.000942&0.70$\pm$0.04& 0.958$\pm$0.000&0.0024$\pm$0.0016& 796 \\
5670-5686 &0.104081$\pm$0.000949&0.70$\pm$0.04& 0.958$\pm$0.000&0.0020$\pm$0.0016& 779 \\
5655-5670 &0.103933$\pm$0.000905&0.70$\pm$0.04& 0.960$\pm$0.000&-0.0020$\pm$0.0016& 801 \\
    \end{tabular}
  \end{center}
\end{table*}

\begin{table*}
  \caption{KOSMOS Final Model Parameters (II)}
  \label{tab:finalpars2}
  \begin{center}
    \leavevmode
    \begin{tabular}{c c c c c c } \hline \hline
Wavelength (A) & R$_{p}$/R$_{s}$ & u$_{1}$ & a$_{0}$ & a$_{1}$ & $\sigma_{res}$ (ppm)  \\
\hline
5639-5655 &0.104094$\pm$0.000967&0.71$\pm$0.04& 0.959$\pm$0.001&-0.0004$\pm$0.0017& 833 \\
5624-5639 &0.103983$\pm$0.001024&0.71$\pm$0.04& 0.960$\pm$0.001&-0.0004$\pm$0.0017& 855 \\
5608-5624 &0.103929$\pm$0.001255&0.71$\pm$0.05& 0.958$\pm$0.001&0.0079$\pm$0.0021& 861 \\
5593-5608 &0.104014$\pm$0.001103&0.72$\pm$0.04& 0.960$\pm$0.001&-0.0024$\pm$0.0019& 799 \\
5578-5593 &0.104133$\pm$0.000996&0.72$\pm$0.04& 0.960$\pm$0.001&0.0001$\pm$0.0017& 786 \\
5562-5578 &0.103605$\pm$0.001056&0.72$\pm$0.04& 0.959$\pm$0.001&0.0059$\pm$0.0019& 843 \\
5547-5562 &0.103690$\pm$0.000915&0.72$\pm$0.03& 0.960$\pm$0.001&0.0046$\pm$0.0016& 800 \\
5531-5547 &0.103654$\pm$0.000926&0.72$\pm$0.03& 0.962$\pm$0.000&0.0027$\pm$0.0016& 831 \\
5516-5531 &0.103758$\pm$0.001074&0.72$\pm$0.04& 0.960$\pm$0.001&0.0032$\pm$0.0018& 845 \\
5500-5516 &0.103802$\pm$0.000914&0.72$\pm$0.04& 0.959$\pm$0.000&0.0037$\pm$0.0016& 781 \\
5485-5500 &0.104148$\pm$0.000965&0.74$\pm$0.04& 0.956$\pm$0.000&0.0068$\pm$0.0016& 815 \\
5470-5485 &0.103887$\pm$0.000885&0.72$\pm$0.04& 0.957$\pm$0.000&0.0046$\pm$0.0014& 893 \\
5454-5470 &0.103850$\pm$0.000963&0.74$\pm$0.04& 0.956$\pm$0.000&0.0064$\pm$0.0016& 827 \\
5439-5454 &0.103816$\pm$0.000911&0.74$\pm$0.04& 0.956$\pm$0.000&-0.0007$\pm$0.0015& 869 \\
5423-5439 &0.104098$\pm$0.000998&0.74$\pm$0.04& 0.955$\pm$0.001&-0.0004$\pm$0.0017& 870 \\
5408-5423 &0.103452$\pm$0.001074&0.72$\pm$0.04& 0.953$\pm$0.001&-0.0042$\pm$0.0018& 913 \\
5392-5408 &0.103725$\pm$0.001055&0.72$\pm$0.04& 0.950$\pm$0.001&0.0066$\pm$0.0018& 886 \\
5377-5392 &0.103753$\pm$0.001019&0.72$\pm$0.04& 0.948$\pm$0.001&0.0069$\pm$0.0018& 861 \\
5361-5377 &0.103740$\pm$0.000980&0.74$\pm$0.04& 0.952$\pm$0.001&0.0024$\pm$0.0016& 811 \\
5346-5361 &0.103821$\pm$0.000962&0.74$\pm$0.04& 0.952$\pm$0.001&0.0018$\pm$0.0016& 812 \\
5114-5346 &0.103906$\pm$0.000501&0.75$\pm$0.02& 0.942$\pm$0.000&0.0013$\pm$0.0008& 564 \\
4883-5114 &0.103907$\pm$0.000488&0.80$\pm$0.02& 0.941$\pm$0.000&0.0020$\pm$0.0008& 585 \\
4651-4883 &0.103833$\pm$0.000558&0.81$\pm$0.02& 0.934$\pm$0.000&0.0018$\pm$0.0009& 637 \\
4419-4651 &0.103620$\pm$0.000551&0.86$\pm$0.02& 0.933$\pm$0.000&0.0001$\pm$0.0009& 669 \\
4188-4419 &0.104013$\pm$0.000744&0.87$\pm$0.03& 0.916$\pm$0.000&0.0020$\pm$0.0012& 647 \\
3956-4188 &0.104116$\pm$0.000898&0.90$\pm$0.03& 0.894$\pm$0.000&0.0012$\pm$0.0014& 737 \\
3724-3956 &0.104188$\pm$0.000899&0.86$\pm$0.03& 0.873$\pm$0.000&-0.0111$\pm$0.0014& 842 \\
    \end{tabular}
  \end{center}
\end{table*}

\end{document}